\documentclass[12pt,preprint]{aastex}
\def\CIVdblt{{\rm C}\kern 0.1em{\sc iv}~$\lambda\lambda 1548, 1550$}
\def\MgIIdblt{{\rm Mg}\kern 0.1em{\sc ii}~$\lambda\lambda 2796, 2803$}
\def\NVdblt{{\rm N}\kern 0.1em{\sc v}~$\lambda\lambda 1238, 1242$}  
\def\OVIdblt{{\rm O}\kern 0.1em{\sc vi}~$\lambda\lambda 1031, 1037$}
\def\SiIVdblt{{\rm Si}\kern 0.1em{\sc iv}~$\lambda\lambda1394, 1403$}
\def\AlIIIdblt{{\rm Al}\kern 0.1em{\sc iii}~$\lambda\lambda1855,1863$}

\def\AlII{\hbox{{\rm Al}\kern 0.1em{\sc ii}}}
\def\AlIII{{\hbox{\rm Al}\kern 0.1em{\sc iii}}}
\def\CaI{\hbox{{\rm Ca}\kern 0.1em{\sc i}}}
\def\CaII{\hbox{{\rm Ca}\kern 0.1em{\sc ii}}}
\def\CrII{\hbox{{\rm Cr}\kern 0.1em{\sc ii}}}
\def\CII{\hbox{{\rm C}\kern 0.1em{\sc ii}}}
\def\CIII{\hbox{{\rm C}\kern 0.1em{\sc iii}}}
\def\CIV{\hbox{{\rm C}\kern 0.1em{\sc iv}}}
\def\CV{\hbox{{\rm C}\kern 0.1em{\sc v}}}
\def\HI{\hbox{{\rm H}\kern 0.1em{\sc i}}}
\def\HII{\hbox{{\rm H}\kern 0.1em{\sc ii}}}
\def\Lya{\hbox{{\rm Ly}\kern 0.1em$\alpha$}}
\def\Lyb{\hbox{{\rm Ly}\kern 0.1em$\beta$}}
\def\Lyg{\hbox{{\rm Ly}\kern 0.1em$\gamma$}}
\def\Lyfive{\hbox{{\rm Ly}\kern 0.1em$5$}}
\def\Lysix{\hbox{{\rm Ly}\kern 0.1em$6$}}
\def\Lyseven{\hbox{{\rm Ly}\kern 0.1em$7$}}
\def\Lyeight{\hbox{{\rm Ly}\kern 0.1em$8$}}
\def\Lynine{\hbox{{\rm Ly}\kern 0.1em$9$}}
\def\Lyten{\hbox{{\rm Ly}\kern 0.1em$10$}}
\def\HeI{\hbox{{\rm He}\kern 0.1em{\sc i}}}
\def\HeII{\hbox{{\rm He}\kern 0.1em{\sc ii}}}
\def\FeI{\hbox{{\rm Fe}\kern 0.1em{\sc i}}}
\def\FeII{\hbox{{\rm Fe}\kern 0.1em{\sc ii}}}
\def\FeIII{\hbox{{\rm Fe}\kern 0.1em{\sc iii}}}
\def\MnII{\hbox{{\rm Mn}\kern 0.1em{\sc ii}}}
\def\MgI{\hbox{{\rm Mg}\kern 0.1em{\sc i}}}
\def\MgII{\hbox{{\rm Mg}\kern 0.1em{\sc ii}}}
\def\MgIII{\hbox{{\rm Mg}\kern 0.1em{\sc iii}}}
\def\MgIV{\hbox{{\rm Mg}\kern 0.1em{\sc iv}}}
\def\NaI{\hbox{{\rm Na}\kern 0.1em{\sc i}}}
\def\NV{\hbox{{\rm N}\kern 0.1em{\sc v}}}
\def\NII{\hbox{{\rm N}\kern 0.1em{\sc ii}}}
\def\NIII{\hbox{{\rm N}\kern 0.1em{\sc iii}}}
\def\OVI{\hbox{{\rm O}\kern 0.1em{\sc vi}}}
\def\OII{\hbox{[{\rm O}\kern 0.1em{\sc ii}]}}
\def\SiII{\hbox{{\rm Si}\kern 0.1em{\sc ii}}}
\def\SiIII{\hbox{{\rm Si}\kern 0.1em{\sc iii}}}
\def\SiIV{\hbox{{\rm Si}\kern 0.1em{\sc iv}}}
\def\SII{\hbox{{\rm S}\kern 0.1em{\sc ii}}}
\def\SIII{\hbox{{\rm S}\kern 0.1em{\sc iii}}}
\def\SIV{\hbox{{\rm S}\kern 0.1em{\sc iv}}}
\def\TiII{\hbox{{\rm Ti}\kern 0.1em{\sc ii}}}
\def\ZnII{\hbox{{\rm Zn}\kern 0.1em{\sc ii}}}
\newcommand{\kms}{\hbox{km~s$^{-1}$}}

\def\kms{\hbox{km~s$^{-1}$}}      
\def\cm2{\hbox{cm$^{-2}$}}

\def\etal{et~al.\ }

\begin{document}
\title{A Survey of Weak {\MgII} Absorbers at $0.4 < z < 2.4$\footnotemark[1]}

\footnotetext[1]{Based on public data obtained from the ESO archive of observations done using the UVES spectrograph at the VLT, Paranal, Chile.}

\author{Anand~Narayanan\altaffilmark{2},Toru~Misawa\altaffilmark{2}, Jane~C.~Charlton\altaffilmark{2}, and Tae-Sun Kim\altaffilmark{3}}

\altaffiltext{2}{Department of Astronomy and Astrophysics, The Pennsylvania State University, University Park, PA 16802, {\it anand, misawa, charlton@astro.psu.edu}}

\altaffiltext{3}{Astrophysiakalisches Institut Postdam, An der Sternwarte 16, 14482 Potsdam, Germany, {\it tkim@aip.de}}

\begin{abstract}
We present results from a survey of weak {\MgII} absorbers in the VLT/UVES spectra of $81$~QSOs obtained from the ESO archive. In this survey, we identified 112 weak {\MgII} systems within the redshift interval $0.4 < z < 2.4$ with $86$~\% completeness down to a rest-frame equivalent width of $W_r(2796) = 0.02$~{\AA}, covering a cumulative redshift path length of $\Delta$Z~$\sim 77.3$. From this sample, we estimate that the number of weak absorbers per unit redshift ($dN/dz$) increases from $1.06~{\pm}~0.04$ at $<z>=1.9$ to $1.76~{\pm}~0.08$ at $<z>=1.2$ and thereafter decreases to $1.51~{\pm}~0.09$ at $<z>=0.9$ and $1.06~{\pm}~0.10$ at $<z>=0.6$. Thus we find evidence for an evolution in the population of weak {\MgII} absorbers, with their number density peaking at $z=1.2$. We also determine the equivalent width distribution of weak systems at $<z>=0.9$ and $<z>=1.9$. At $0.4 < z < 1.4$, there is evidence for a turnover from a powerlaw of the form $n(W_r)~ \propto~ W_r^{-1.04}$ at $W_r(2796) < 0.1$~{\AA}. This turnover is more extreme at $1.4 < z < 2.4$, where the equivalent width distribution is close to an extrapolation of the exponential distribution function found for strong {\MgII} absorbers. Based on these results, we discuss the possibility that some fraction of weak {\MgII} absorbers, particularly single cloud systems, are related to satellite clouds surrounding strong {\MgII} systems. These structures could also be analogs to Milky Way high velocity clouds. In this context, the paucity of high redshift weak {\MgII} absorbers is caused by a lack of {\it isolated} accreting clouds on to galaxies during that epoch. 
\end{abstract}

\section{INTRODUCTION}
\label{sec:1}

Weak {\MgII} absorbers (those with {\MgII}~$\lambda 2796$~{\AA}
rest frame equivalent width $W_r(2796)<0.3$~{\AA}) represent a 
population or populations distinct
from the stronger {\MgII} absorbers which are directly associated
with luminous galaxies (L $> 0.05$L$^*$).  This conclusion is based partly
upon a rapid rise in the equivalent width distribution of $W_r(2796)$ at
values below $0.3$~{\AA} \citep{weak1,nestor06}.
It is also partly based upon the excess of single-cloud weak {\MgII}
absorbers, over that expected from the Poisson distribution of number
of clouds per system found for strong {\MgII} absorbers \citep{weak2}.  
The single-cloud weak {\MgII} absorbers comprise $\sim 2/3$
of the weak {\MgII} absorber population at $0.4 < z < 1.4$, with the
remainder having multiple clouds in {\MgII} absorption.

The single-cloud weak {\MgII} absorbers tend to have
metallicities $>0.1$ times the solar value, and in some cases greater
than the solar value \citep{weak2,weak1634}. Though data are limited, 
it is clear that most single-cloud weak
{\MgII} absorbers are not produced by lines of sight very close to luminous
galaxies, though most are found at impact parameters of $30$-$100
h^{-1}$~kpc \citep{cwc05,milni06}.  Thus their high metallicities
are surprising.  Furthermore, the large ratio of {\FeII} to {\MgII}
column density in some weak {\MgII} absorbers indicates that ``in situ''
star formation is responsible for their enrichment \citep{weak2}.

Photoionization modeling of the single-cloud weak {\MgII} absorbers
has established the existence of two phases, a high density region
that is $1-100$~pc thick and produces narrow ($\sim$few {\kms}) low
ionization lines, and a kiloparsec scale, lower density region that
produces somewhat broader, high ionization lines.  There are often
additional, similar low density regions within tens of {\kms} of the
one that is aligned with the {\MgII} absorption.  \citet{milni06} 
argue that filamentary and sheetlike geometries are
required for the single-cloud, weak {\MgII} absorbers, based on a
census of the absorber populations at $0 < z < 1$, and discussed
possible origins in satellite dwarf galaxies, in failed dwarf galaxies,
or in the analogs to Milky Way high velocity clouds.  The earlier work
of \citet{weak2} considered Population III star clusters, star
clusters in dwarf galaxies, and fragments in Type Ia supernovae shells
as possible sites for production of weak {\MgII} absorbers.  Most
recently, \citet{LC06b} have argued that the close alignment in
velocity of the {\MgII} and {\CIV} absorption is also suggestive of a
layered structure such as expected for supernova remnants or for high
velocity clouds sweeping through a hot corona.

Single-cloud weak {\MgII} absorbers have possible implications for
star formation in dwarf galaxies and in the intergalactic medium, and
for tracking the populations of dwarf galaxies and/or high velocity
clouds to high redshifts.  For example, \citet{LCT06a} noted
that the peak at $z\sim1$ of the star formation rate in dwarf galaxies
may be related to the evolution of the weak {\MgII} absorbers.  To 
understand the relative importance of the processes that produce
weak {\MgII} absorption, it is crucial to have accurate measures of
the evolution of their number densities.

Multiple-cloud weak {\MgII} absorbers may also be important as a tool
to trace evolution of dwarf galaxies and other metal-rich gas too
faint to see at high redshifts.  Because of their abundance, the dwarf
galaxy population should present a significant cross section for
absorption, yet so far their absorption signatures have been hard to
recognize. Although some of the multiple-cloud weak {\MgII} absorbers
are surely an extension of the strong {\MgII} absorber population,
others are kinematically compact, and are possibly related to dwarf
galaxies \citep{zonak,masiero05,ding05}. It is of interest to have a 
survey of weak {\MgII} absorbers large enough to separately consider 
the evolution of multiple-cloud weak {\MgII} absorbers.

There have been three comprehensive surveys for weak {\MgII}
absorbers, each focused on a different redshift regime. 
\citet[hereafter CRCV99]{weak1} report on a survey for weak {\MgII}
systems in the interval $0.4 < z < 1.4$, \citet{anand05}
covered the range $0 < z < 0.3$, and more recently 
\citet[hereafter LCT06]{LCT06a} discovered weak systems in the redshift
interval $1.4 < z < 2.4$. These studies followed the earlier, smaller
surveys by \citet{womble} and \citet{tripp}, who first established
that the equivalent width distribution of {\MgII} absorbers continues
to rise below $W_r(2796) =0.3$~{\AA}.  The number density ($dN/dz$)
constraints from the surveys collectively demonstrate an evolution in
the absorber population over the redshift interval $0 \leq z \leq
2.4$, comprising the last $\sim 10$~Gyr history of the universe.

For their survey of weak {\MgII} systems, CRCV99 searched a redshift
pathlength of $\Delta Z = 17.2$ in the HIRES/Keck spectra of 26
QSOs. Thirty weak {\MgII} systems were identified in the interval $0.4
< z < 1.4$ in that survey, which was $80$\% complete down to an
rest frame equivalent width sensitivity limit of $W_r(2796) =
0.02$~{\AA}. From the weak {\MgII} systems identified, they estimated
a redshift path density $dN/dz = 1.74 \pm 0.10$ for $<z> = 0.9$, and
for $0.02 \leq W_r(2796) < 0.3$~{\AA}.  Later, using STIS/HST UV
echelle spectra of 20 quasars, \citet{anand05} found that
analogs to weak {\MgII} absorbers at $z \sim 1$ also exist in the
present universe. From the six systems detected in a redshift
pathlength of $\Delta Z=5.3$ within the redshift window $0 < z <
0.3$, a $dN/dz$ of $1.00 \pm 0.20$ was estimated for $<z> = 0.15$.
LCT06 presents the most recent survey for weak {\MgII} absorbers. From
a data set of 18 QSOs, observed using the UVES/VLT, a total of 9 weak
systems were found over a redshift path of $\Delta Z =8.5$ in the interval
$1.4 < z < 2.4$, yielding a $dN/dz = 1.02 \pm 0.12$ for $<z> = 1.9$. 
That survey was $100\%$ complete down to a rest frame equivalent width 
of $W_r(2796) = 0.02$~{\AA}.

In order to interpret the apparent evolution in the $dN/dz$ of weak
{\MgII} absorbers it is necessary to consider the effect of the
changing extragalactic background radiation (EBR).  The EBR is known
to diminish in intensity by $\sim 0.5$~dex from $z=2$ to $z=1$, and by
$\sim 1$~dex from $z \sim 1$ to $z \sim 0$ \citep{hm96,hm01}.  This 
changing EBR will have an effect on, what might otherwise
be a static population of absorbers, due to a change in the balance
between high and low ionization gas.  However, what we would predict
from the EBR evolution would be an increase in $dN/dz$ from $z \sim 2$
to $z \sim 0$.  This makes the smaller observed $dN/dz$ at $<z>=0.15$
quite significant, in that it implies a real decrease in the population
from $<z>=0.9$ to $<z>=0.15$ \citep{anand05}.  Similarly,
LCT06 found that the increase in $dN/dz$ from $<z>=1.9$
to $<z>=0.9$ was significantly larger than that predicted from the
effect of the changing EBR (and the expected cosmological evolution).
Thus, in light of the results from the three surveys, it can be argued that
there has been a slow build up of weak systems from high redshift,
with their number density reaching a peak at $z \sim 1$, and
subsequently evolving away until the present time. 

The goal of the present study is to determine more precisely how
$dN/dz$ evolves at $z>1$.  The LCT06 survey identified an overall
trend in number density evolution, but was limited by small sample
size. Our sample covers $\sim 4.5$ times more lines of sight than
LCT06. This will allow us to constrain $dN/dz$ for smaller redshift bins
in order to measure a peak redshift for the incidence of weak {\MgII}
absorption.  A larger sample will also allow us to look separately
at the evolution of the single-cloud and multiple-cloud weak {\MgII}
absorption, which is important because they are likely to originate
in different types of structures.  Finally, we will examine the
equivalent width distribution for weak {\MgII} absorbers, and
consider its evolution.

In \S~\ref{sec:2} we describe the VLT/UVES dataset and outline our procedures
for reducing the spectra and for searching for weak {\MgII} doublets.
\S~\ref{sec:3} presents the formal results of our survey, including the
redshift path density for $W_r(2796)>0.02$~{\AA} absorbers at $0.4 < z < 2.4$,
separates this into single-cloud and multiple-cloud weak {\MgII} absorbers,
and presents the equivalent width distributions at $<z>=0.9$ and $<z>=1.9$.
A summary and discussion is given in the final section of the paper.

\section{DATA \& SURVEY METHOD}
\label{sec:2}

\subsection{UVES/VLT Archive Data}
\label{sec:2.1}

Our sample of 81 quasar spectra used for the survey was retrieved from
the ESO archive.  Since there is no comprehensive method to find
all quasar spectra in the archive, we searched for programs
with titles and abstracts that seemed relevant.
We then retrieved all $R \sim 45,000$ spectra made available before
June 2006.  The spectra were obtained to
facilitate various studies of stronger metal-line absorbers and of the
{\Lya} forest, but in no case should there be a particular bias toward
or against weak {\MgII} systems.  We eliminated several spectra which
had $S/N < 30$ pixel$^{-1}$ over their full wavelength coverage because
those would compromise our survey completeness at small equivalent
widths.

The reduction and wavelength calibration of the echelle data were
carried out using the ESO provided MIDAS pipeline. To enhance the
$S/N$ ratio of the spectra, all available observations of a particular
target were included in the reduction. The reduced one dimensional
spectra were vacuum--heliocentric velocity corrected and rebinned to
$0.03$~{\AA}, corresponding to the pixel width in the blue part of the spectrum.  
The different exposures for a particular target were each scaled by
the median ratio of counts from the exposure with the best $S/N$ to
the counts from that exposure itself.  This puts all the exposures
on the same relative flux scale. The
scaled spectra were then co-added, weighting by the $S/N$
corresponding to each pixel.  Continuum fitting was done on the
reduced spectra using the IRAF{\footnote {IRAF is distributed by the
National Optical Astronomy Observatories, which are operated by AURA,
Inc., under cooperative agreement with NSF}} SFIT procedure. The
spectra were then normalized by the continuum fit.

Table~\ref{tab:1} provides a detailed list of the quasars that were
used for this survey. The UVES offers a large wavelength coverage,
from $3000$~{\AA} to $1$~$\mu$m, thus spectra often include many
different chemical transitions for an absorption system. However, the
wavelength coverage available for individual quasar spectra varied,
based on the choice of cross-disperser settings. Combining exposures
from various settings therefore sometimes resulted in gaps in
wavelength coverage. In addition to wavelength gaps, we systematically
excluded the following path lengths from our formal search: (1)
wavelength regions blueward of the {\Lya} emission line, as they are
strongly affected by forest lines; (2) wavelength regions that are
within 5000 {\kms} of {\MgIIdblt} emission corresponding to the
redshift of the quasar, as any absorption line within this regime has
a higher probability of being intrinsic; and (3) regions of the
spectrum that are polluted by various atmospheric absorption features,
including the A and B absorption bands from atmospheric oxygen. The
elimination of wavelength regions that are affected by the telluric
lines was complicated, since some spectra were affected more than
others.  This prohibited us from eliminating equal redshift paths from
all quasars, since by doing so we would have discarded wavelength
regions that are suitable for searching weak lines.

\begin{figure*}
\figurenum{1}
\epsscale{0.8}
\rotatebox{90}{\plotone{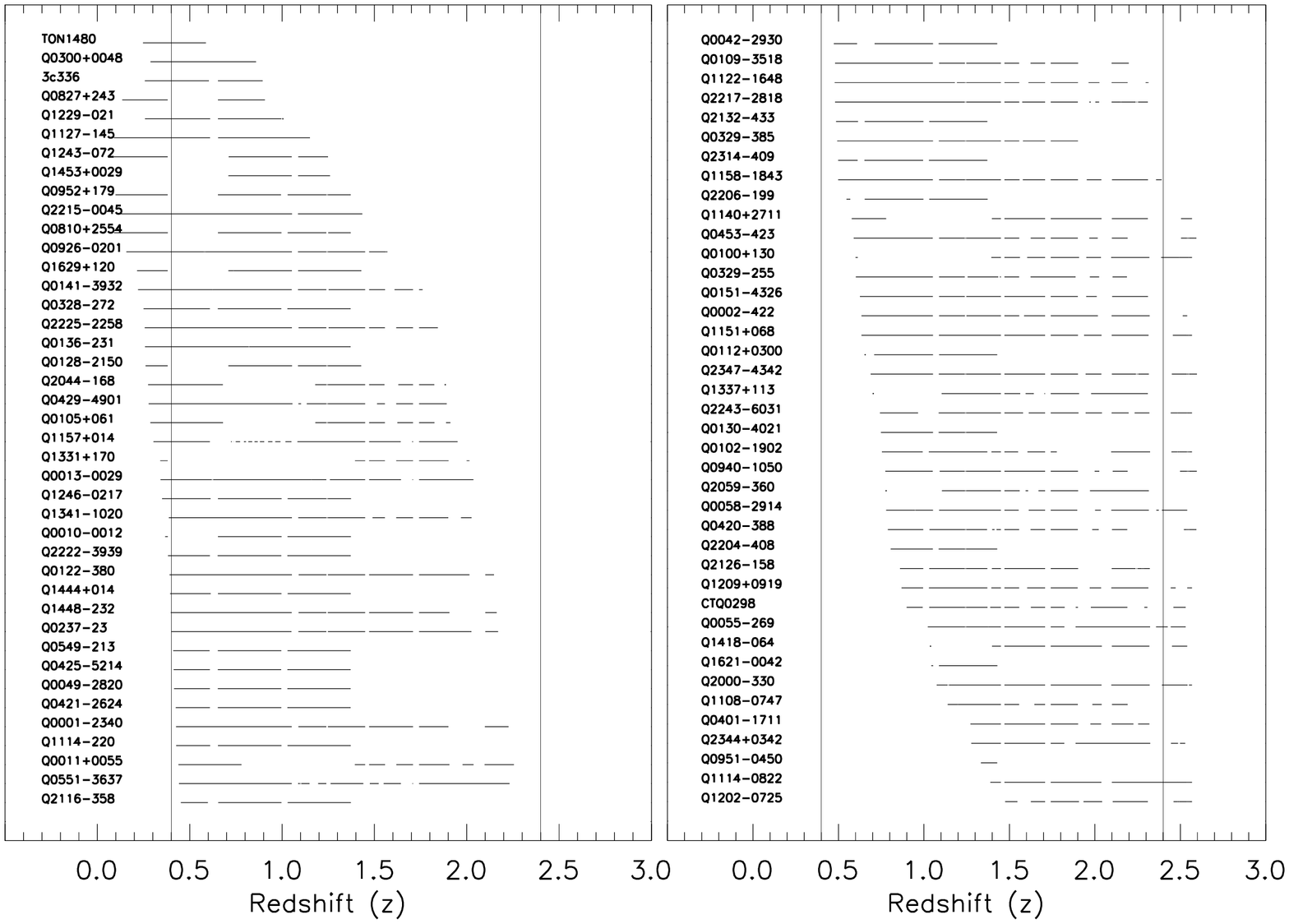}}
\protect
\caption{The path length that was available in each quasar spectrum. The quasars are arranged from top to bottom and from left to right panel in increasing order of emission redshift (refer Table 1). The two vertical lines at $z=0.4$ and $z=2.4$ mark the limiting boundaries of our survey. We confined the lower redshift limit of our survey to $z=0.4$ since the coverage from our sample dropped significantly below that redshift.}
\label{fig:1}
\end{figure*}

Figure~\ref{fig:1} illustrates the redshift path length that was available in
each quasar spectrum for an {\MgIIdblt} search. The wavelength regions
that were thickly contaminated with telluric lines (typically at
$\lambda > 8000$~{\AA}) were eliminated if the observed equivalent
width of a significant number of those lines were equal to or greater than
the $W_r(2796) = 0.02$~{\AA} (the lower equivalent limit of our
survey). The possibility of chance alignment between atmospheric
lines, in most cases, was resolved by confirmation with associated absorption
features that were covered and detected. This confirmation procedure
was most feasible for {\MgIIdblt} at high redshifts, where
additional metal lines for the system (such as {\FeII}~$\lambda$~$2600$,
{\CIVdblt}, {\CII}~$\lambda$~$1335$, etc.) are covered in the blue portion
of the wavelength coverage.

\subsection{Survey Method}
\label{sec:2.2}

In searching for {\MgII} systems in the included redshift path of each
quasar spectrum, we first assumed every absorption line detected at an
equivalent width limit of $5\sigma$ as the {\MgII}~$\lambda$~$2796$
line of a possible {\MgII} doublet. A candidate {\MgII} system was
considered if there was at least a $2.5\sigma$ detection of the
corresponding $\lambda$~$2803$ line for the same redshift. The lines
of the doublet were also visually inspected for comparable profile
shapes and for a doublet ratio between 1:1 and 2:1. The detected
system was considered to be a weak {\MgII} absorber if the measured
rest-frame equivalent width, $W_r(2796)$, was less than
$0.3$~{\AA}. To further confirm the detection, we also looked for
associated metal lines (e.g. {\FeII}, {\MgI}, {\CIV}, {\SiIV}, etc.)
and {\Lya} that were covered and likely to be detected for weak
systems.  Weak {\MgII} doublets that were found within $500$~{\kms} of
each other were taken as part of the same absorbing system, and are
therefore classified as one multiple cloud system.  In one case, the
system at $z=1.0446$ towards Q~$2314-409$, grouping together two weak
components, separated by $134$~{\kms}, in this way led to a
classification as a strong {\MgII} absorber, and thus exclusion from
our survey.  Finally, as in CRCV99, in order to be considered as a
separate system, a weak {\MgII} absorber must be at least
$1000$~{\kms} from any strong {\MgII} absorption.  

Using the $81$ QSO lines of sight, we detected $116$ weak {\MgII} system in total.
Out of this, $112$ systems are within the redshift interval $0.4 < z < 2.4$.
Our redshift coverage drops significantly at $z < 0.4$, and therefore
we limit our survey to within $0.4 < z < 2.4$. This further helps to directly
compare our results to the preceding surveys of CRCV99 and LCT06, which were also
confined to the same redshift interval.
Table~\ref{tab:2} provides the complete sample of weak {\MgII} absorbers that we 
identified, and figures~\ref{fig:2a}--~\ref{fig:2e} illustrate the 
{\MgIIdblt} absorption profiles of a few example cases from the systems that we identified. ({\it The absorption profile of all systems identified in our survey will be published in the online version of the journal}).

In our doublet search, a certain number of candidate
{\MgII}~$\lambda$~$2796$ features, with detections at the position of
the corresponding $2803$ turned out to be chance alignments. To
illustrate that these cases are well-understood and do not lead
to significant uncertainty in our sample, we describe those
instances:

\begin{enumerate}

\item{In the spectrum of Q~$1122-1648$, a candidate weak {\MgII} doublet was
detected at redshift $z = 0.5109$. Visual inspection showed that the
profile shapes of the doublet lines were inconsistent with each
other. The {\MgII}~$\lambda$~$2796$ feature was later identified as
the {\CIV}~$\lambda$~$1551$ line of a {\CIVdblt} from an absorption
system at $z = 1.7244$, further confirmed by the presence of {\Lya}
at $\sim 3311$~{\AA}.}

\item{In the spectrum of Q~$1158-1843$, a candidate weak {\MgII} doublet
was detected at $z = 1.1700$ for which the {\CIVdblt} was covered,
but not detected. Subsequently, the {\MgII}~$\lambda$~$2803$ feature
was identified as the {\AlIII}~$\lambda$~$1863$ line of the
{\AlIIIdblt} doublet at $z = 2.2660$ for which associated {\Lya}, {\CIVdblt},
{\SiIVdblt}, {\CII}~$\lambda$~$1335$, {\SiII}~$\lambda$~$1260$,
{\SiII}~$\lambda$~$1527$, etc., were also detected.}

\item{A possible weak {\MgIIdblt} detection was found at $z = 1.2400$
along the line of sight to Q~$2314-409$ and was ruled out as
chance alignment because of significant mismatch between profile shapes. 
Metal lines, such as {\CIV}, {\SiIV}, {\CII} or {\SiII}
and {\Lya}, for this prospective system were not covered in the
spectrum.}

\item{The candidate {\MgIIdblt} absorption feature at $z = 1.2433$ in the
spectrum of Q~$2225-2258$ did not have any high ionization {\CIV} or
{\SiIV} detected. What was identified as the {\MgII}~$\lambda$~$2796$
feature was subsequently identified as the {\FeII}~$\lambda$~$2600$
absorption line for the weak {\MgII} system at $z = 1.4126$.}

\item{The possible {\MgIIdblt} detection at $z = 1.8271$ in Q~$1202-0725$ 
was dismissed
from consideration as a weak system since the {\MgII}~$\lambda$~$2796$
and $2803$ profile shapes were not consistent with expectations for a
doublet.  The detection of {\CIV} and {\SiIV} for that redshift could
not be confirmed, since those features would have been located in the
region of the spectra that was densely populated by forest
lines. Other low ionization transitions, such as
{\SiII}~$\lambda$~$1260$ or {\CII}~$\lambda$~$1335$, did not fall
within the wavelength coverage of the spectrum.}

\item{The candidate $z = 2.2124$ {\MgII} system in Q~$2000-330$ was ruled out.
It was considered a very unlikely candidate because of profile shapes 
between the members of the doublet not being consistent with each other, 
and also because of the equivalent width ratio, $W_r(2796)/W_r(2803)$, being 
significantly less than 1. {\SiIV} and {\CIV} would have been in the 
region of the spectrum that was densly contaminated by the forest, and 
therefore could not be identified.}

\end{enumerate}

To facilitate comparison with previous surveys by CRCV99 and LCT06, we
confine the equivalent width range of our survey to $0.02 \leq W_r(2796) <
0.3$~{\AA}. Of the $116$ weak {\MgII} systems detected in our survey,
three were measured to have $W_r(2796) < 0.02$~{\AA} (refer table~\ref{tab:2})
and they are excluded from our $dN/dz$ calculations. However, these weaker 
systems are extremely important to understand whether there is a turnover
in the equivalent width distribution below some limiting value.
Similarly, our redshift coverage drops off dramatically  below
$z=0.4$, with only four systems found, thus we limit our survey
to the range $0.4 < z < 2.4$.

\section{SURVEY COMPLETENESS AND REDSHIFT NUMBER DENSITY}
\label{sec:3}

\subsection{Survey Completeness}
\label{sec:3.1}

The survey completeness is dependent on the detection sensitivity at
different equivalent widths over the redshift path length of the
survey. The detection sensitivity, defined by the likelihood of
detecting a weak {\MgII} doublet along a given path length to a
quasar, is dependent on the quality of the spectrum and also on the
strength of the absorption feature. The survey completeness was
calculated using the formalism given by \citet{ss92} and \citet{lanzetta87}. 
Figure~\ref{fig:3} shows the completeness of our survey at different {\MgII}~$\lambda$2796
equivalent width limits.  We find that our survey is 86~$\%$
complete at the limiting equivalent width of $W_r(2796) = 0.02$~{\AA},
for the redshift path length $0.4 < z < 2.4$. In comparison, the
CRCV99 survey was 80~$\%$ complete for $0.4 < z < 1.4$, and LCT06 was
100~$\%$ complete for $1.4 < z < 2.4$ and for the same equivalent
width limit. The higher completeness of LCT06 is due to their sample 
of $18$~QSOs having better $S/N$. Table~\ref{tab:2} also lists the total redshift path length
$\Delta$Z over which each system discovered in our survey could have
been detected from our sample of lines of sight.

\subsection{Redshift Number Density}
\label{sec:3.2}

The redshift number density, $dN/dz$, of weak {\MgII} absorbers
is calculated using the expression:

\begin{equation}	
\frac{dN}{dz} = \sum_{i}^{N_{sys}} [Z(W_i,R_i)]^{-1},
\end{equation}

summing over all systems, where $Z(W_i,R_i)$ is the cumulative
redshift pathlength covered in the total survey
at rest frame equivalent width $W_i$ for the $i$-th {\MgII} doublet with
doublet ratio $R_i$.  This expression therefore includes small corrections
for incompleteness at small $W_r(2796)$.
Similarly, the variance in $dN/dz$ is given by 
\begin{equation}
\sigma_{dN/dz}^{2} = \sum_{i}^{N_{sys}}[Z(W_i,R_i)]^{-2}.
\end{equation}

Including all quasars, we find a total redshift pathlength $\Delta Z \sim 77.3$
for this survey over the range $0.4 < z < 2.4$.  Of this redshift pathlength,
$\Delta Z \sim 50.7$ is in the lower redshift regime ($0.4 < z < 1.4$), 
as compared to $\Delta Z \sim 17.2$ for CRCV99. Our coverage in the higher 
redshift regime ($1.4 < z < 2.4$) is $\Delta Z \sim 26.6$ as compared to 
$\Delta Z = 8.5$ for LCT06.

For systems within the equivalent width range $0.02 \leq W_r(2796) < 0.3$~{\AA}, the number densities for the various redshifts intervals (chosen for comparison with previous surveys) are listed in table~\ref{tab:3}.

Our larger survey size enabled the error bars in these
estimations to be constrained to values smaller than those for
the previous surveys. Figure~\ref{fig:4} shows the $dN/dz$ values
for the various redshift ranges. Our $dN/dz$ estimate is, in general,
consistent with the results from previous surveys. For the redshift
bin $0.7 < z < 1.0$, the constraints from CRCV99 and LCT06 
differed by more than $2$~$\sigma$. Our result for this redshift bin is
closer to the measurement from CRCV99, suggesting that the LCT06
point was off because of statistical fluctuations due to the small 
sample. It is important to note that our $1.4 \leq z \leq 2.4$ 
datapoint is in agreement with the earlier survey of LCT06.  This is 
an important verification that we are correcting our numbers 
appropriately to account for the
fact that the spectra in our larger sample were, on average, of
slightly lower quality than those surveyed by LCT06.

In figure~\ref{fig:5}, we focus on just the present VLT/UVES sample, and examine evolution of weak absorbers within the redshift interval $1.4 < z < 2.4$
more sequentially, with smaller redshift bins ($\Delta$z $\sim
0.3$).  We can now see that not only is there a drop in the number
density of weak {\MgII} absorbers at $z>1.4$, but it appears to be
a steady drop.  There is a distinct peak in $dN/dz$ in the bin centered
at $z=1.2$.

Classifying the absorption systems in our sample as single
cloud (a single kinematic component) and/or multiple cloud (with more
than one kinematic components), we also calculated the redshift number
densities of both classes separately for the various redshift
bins. These are shown in also shown in figure~\ref{fig:5}.  For consideration of this issue, even our larger sample is suffering from small number statistics.
However, we see that both single-cloud and multiple-cloud weak {\MgII}
absorbers do appear to exhibit a rise and then a fall in their number
densities between $z=2.4$ and $z=0.4$.

\subsection{Equivalent Width Distribution}
\label{sec:3.3}

The equivalent width distribution of {\MgII} systems is typically
parameterized by fitting the data using either an exponential
relationship of the form

\begin{equation}	
\frac{\it{d}n(W)}{\it{d}W} = (N^*/W^*) e^{-W/W^*}
\end{equation}
\noindent  where N$^*$ and $W^*$ are best-fit parameters,
or a power-law relationship of the form 
\begin{equation}
\frac{\it{d}n(W)}{\it{d}W} = CW^{-\delta}
\end{equation}
\noindent where C, a constant, and $\delta$, the power law index, are best-fit parameters.

\subsection{Equivalent Width Distribution at $<z> = 0.9$}
\label{sec:3.3.1}

Using a single power law, with $\delta = 1.04$ and $C=0.54$, CRCV99
were able to produce an acceptable fit to the equivalent width
distribution of both strong and weak systems, with the exception of
the bin centered on the strongest absorbers at $W_r(2796) = 2.2$~{\AA}.
The distribution indicated that, at $<z>=0.9$, there is a drastic increase in the
number of systems toward the weak end of the distribution,
with no indication of turn over in the power law distribution
down to $W_r (2796) = 0.02$~{\AA} (see Fig. 6 of CRCV99).

Figure~\ref{fig:6} shows the distribution function from results based on our
survey for the redshift interval $0.4 < z < 1.4$ and for $W_r(2796)
\geq 0.0165$~{\AA}. The redshift interval and equivalent width lower
limit were selected to be coincident with the values used by CRCV99.
Since the equivalent width distribution is rapidly rising toward small values,
it is critical to make comparisons in the same bins.
For the three equivalent width bins at $W_r(2796) < 0.3$~{\AA}, the
distribution for the bins centered at $0.15$~{\AA} and $0.25$~{\AA}
are consistent with the results from CRCV99 survey, to within
$\sim 1\sigma$. However, our measurement of $n(W_r)$ for the
lowest bin, at $0.06$~{\AA}, is roughly a factor of two less than the
CRCV99 result, a difference of $1.8 \sigma$. Our measurement shows
that there {\it is} a turn-over from the power law equivalent width distribution
suggested in CRCV99, for
$W_r(2796) <0.1$~{\AA}.  We considered the possibility that we
are missing some of the weakest systems in our survey, but we
think this is quite unlikely.  Our survey is $78$~\%
complete at the lower equivalent width limit of $W_r(2796) = 0.0165$~{\AA}, 
whereas the CRCV99 survey, in comparison, is $70$~\% complete at 
that same limit.
We also note that if we use a limiting equivalent width of $0.02$~{\AA}
the discrepancy between the two survey results is at a negligible $1\sigma$ level. 
Thus we confirm that our results at $0.4 < z < 1.4$
are in agreement with CRCV99 in the sense that
weak systems exceed strong systems in number by a factor of $\sim 3:1$.

More recently \citet{nestor05} presented results from a larger
survey of strong {\MgII} absorption systems, identified in the spectra
of $3700$ SDSS quasars.  The equivalent width distribution of their
sample ($0.3 \leq W_r(2796) \leq 5.68$~{\AA}) was fit using the exponential
form described in equation (3), however the fit parameter, $W^*$ and
the resultant normalization $N^*$, were $\sim 1\sigma$ lower than the
parameters derived from the much smaller survey of \citet{ss92}.  
Figure 6 shows the fits from the various parameterizations
for the equivalent width distributions of strong {\MgII} absorbers,
with the more accurate results of \citet{nestor05} shown as the
dashed curve. It is evident that an extrapolation of the exponential
fit to the strong {\MgII} absorbers significantly underestimates the
incidence of weak systems at $<z>=0.9$.

\subsection{Equivalent Width Distribution at $<z> = 1.9$}
\label{sec:3.3.2}

In figure~\ref{fig:7}, we present the equivalent width distribution for weak
{\MgII} absorbers in the range $1.4 < z < 2.4$, and compare to that of
the $0.4 < z < 1.4$ from our VLT/UVES sample.  All low redshift
datapoints are higher than the corresponding high redshift datapoints,
due to the larger overall $dN/dz$ at $<z>=0.9$ than at $<z>=1.9$.  The
plot is log/linear in order to facilitate comparison to the equivalent
width distribution of strong {\MgII} absorbers.  \citet{nestor05}
computed this distribution at redshift $1.311 < z < 2.269$, fitting it
with the parameters $W^* = 0.804$ and $N^* = 1.267$. In Fig. 7, this
function is given as a solid line.  In fact, our datapoints for the
$0.1$--$0.2$~{\AA} and $0.2$--$0.3$~{\AA} bins are consistent with an
extrapolation of the equivalent width distribution for strong {\MgII}
absorbers.  Even the $0.0165$--$0.1$~{\AA} bin is only a factor of two
above the extrapolation.  In contrast, Fig. 7 also shows that at $0.4
< z < 1.4$, there are significantly more weak {\MgII} absorbers (in
all three equivalent width bins) than expected from an extrapolation
of the strong {\MgII} absorber distribution function.  The discrepancy
is more than a factor of ten in the $0.0165$--$0.1$~{\AA} bin.

\section{Summary and Discussion}
\label{sec:4}

We have surveyed the VLT/UVES spectra of 81 quasars to search for weak {\MgII}
absorbers over a redshift path $\Delta Z = 77.3$, in the range
$0.4 < z < 2.4$.  Our survey is 86\% complete at a rest-frame equivalent
width limit $W_r(2796) = 0.02$~{\AA}.  We confirm the result of LCT06
of a declining number density, $dN/dz$ of weak {\MgII} absorbers at $z>1.4$, finding a
peak at $z \sim 1.2$ (see Figure 5).  This general behavior is exhibited separately
for the single and multiple-cloud weak {\MgII} absorbers.  There may be differences
in the evolution of these two classes, but they cannot be distinguished with a
sample of the present size.

At $<z>=0.9$, the equivalent width distribution function for weak {\MgII}
absorbers, shown in figure~\ref{fig:7}, rises substantially above an extrapolation of the exponential
distribution that applies for strong {\MgII} absorbers \citep{nestor05}.
However, at $<z>=1.9$, not only do we see a smaller number of weak {\MgII}
absorbers relative to the expectations from evolution, in figure~\ref{fig:7} we see only a
slight excess over the extrapolation of the strong {\MgII} absorber distribution.

There may not be a very large separate weak {\MgII} absorbers population at $<z>=1.9$,
and at higher redshifts. For example, if we were to extend a linear fit to the four highest
redshift datapoints in figure~\ref{fig:5}, we would predict there would be no weak {\MgII} absorbers
at $z>3$.  Clearly, such an extrapolation is not realistic, since weak {\MgII}
absorption is likely to have multiple causes at any redshift, however it highlights
the fact that there really is a drastic evolution occurring.

LCT06 pointed out a rough coincidence between the peak period of
incidence in weak {\MgII} absorbers (at $z \sim 1$) and the global star formation
rate in the population of dwarf galaxies.  More generally, it seems plausible that 
the evolution in $dN/dz$ of weak {\MgII} absorbers would relate to the rates of
processes that give rise to this absorption.  This remains feasible in view of
the findings of our present survey.  However, a variation of this type of scenario
comes to mind based upon a recent study of the kinematics of strong {\MgII}
absorbers by \citet{mshar06}.  In this new scenario it is not that the
weak {\MgII} absorbers are not being generated at $z>2$.  Instead, these
structures would be evident, at high redshift, as parts of different types
of absorbers, mostly as components of strong {\MgII} absorbers. The basis of 
this suggestion is this hypothesis that there is a three
way connection between weak {\MgII} absorbers, satellite clouds of strong
{\MgII} absorbers, and the extragalactic analogs of the Milky Way high
velocity clouds \citep{mshar06}.  At $z\sim1$ many galaxies exist that are morphologically
and kinematically similar to those in the present epoch \citep{charlton98}.  
Typically, they have a dominant absorbing
component as expected for a galaxy disk, with one or two weaker outlying components
(i.e. satellite clouds) separated by $50$--$300$~{\kms} from the main one.
These satellite clouds look very similar to Milky Way high velocity clouds in
their multiphase absorption properties \citep{fox05,collins05,fox06}.  They
also seem similar to single-cloud weak {\MgII} absorbers, which seem to
have sheetlike or filamentary structures \citep{milni06}, perhaps like
those of Milky Way {\OVI} high velocity clouds.  Furthermore, if weak {\MgII}
absorbers are predominantly found $30$--$100$~kpc from luminous galaxies,
since they have a substantial cross--section relative to the galaxies themselves,
it seems plausible that they are related to high velocity clouds.

\citet{mshar06} find an evolution in the kinematics of strong {\MgII}
absorbers over the same redshift range that we are claiming evolution of
the weak {\MgII} absorbers.  The nature of the evolution is that the strong
{\MgII} systems have a larger number of components at $z\sim2$ than at
$z\sim1$, though their velocity spreads do not change.  These extra components
are very weak, but they act to fill in most of the velocity space spanning
the full range of absorption.  There are no longer separate and distinct
``satellites'', nor is there evidence for single, well-formed galaxies.
In fact, this seems quite analogous to the changes that take place in the
visible morphologies of galaxies from $z\sim2$ to $z\sim1$.  At the higher
redshift galaxies typically have a clump-cluster \citep{elmegreen}
or Tadpole-like morphology, with many separate star-forming regions.
The kinematics of these systems are surely complex and it is likely that
gas is spread through the region.

Finally, returning to the evolution of the weak {\MgII} absorbers that
we have surveyed.  We propose that the absence of them at $z\sim2$ may
be related to a lower probability of passing through just a single
weak {\MgII} absorber.  If the gas that produces {\MgII} absorption is
really so irregularly distributed at $z\sim2$ as suggested by the
strong {\MgII} absorber kinematics, this seems plausible.  It is a
particularly appealing explanation if weak {\MgII} absorbers are the
extragalactic high velocity clouds clustered among the protogalactic
structures in a typical group.  The structures that would produce
single--cloud {\MgII} components and those that would produce
multiple--cloud {\MgII} absorbers may be similar in this respect, in
that both might tend to be kinematically connected at $z\sim2$.
It would be rare to observe an isolated single--cloud weak {\MgII}
absorber because it would be kinematically connected to other
{\MgII} absorbers.  The same could apply for multiple--cloud
weak {\MgII} absorbers if they are also produced by structures that
tend to be concentrated around galaxies.  At $z\sim1$ these
same types of structures form, not necessarily at an increased
rate, but those that do form tend to be more separated from other
absorbing structures for a longer period of time.  This could
produce the peak in the $dN/dz$ distribution of weak {\MgII} absorbers
that we observe at $z\sim1$.  Subsequently, the processes that produce
the structures that produce weak {\MgII} absorption (and perhaps high
velocity clouds as well) may decline in order to give rise to the
declining $dN/dz$ to the present.

Near--IR surveys of weak {\MgII} absorbers at $z>2.4$ will be
needed to determine if the decline found up to this redshift
continues up to higher values.  Furthermore, detailed comparisons of
the physical properties of weak {\MgII} absorbers and the satellite
clouds surrounding strong {\MgII} absorbers.  Finally, comparisons
of the evolution of the ensemble of absorbers to the ensemble of
gas distributions in high redshift groups, both from an
observational and theoretical point of view, is ultimately
needed.

This research was funded by NASA under grants NAG5-6399 and NNG04GE73G
and by the National Science Foundation (NSF) under grant AST-04-07138.
We also acknowledge the ESO archive facility for providing data.


\clearpage

\begin{figure*}
\figurenum{2a}
\epsscale{0.8}
\rotatebox{90}{\plotone{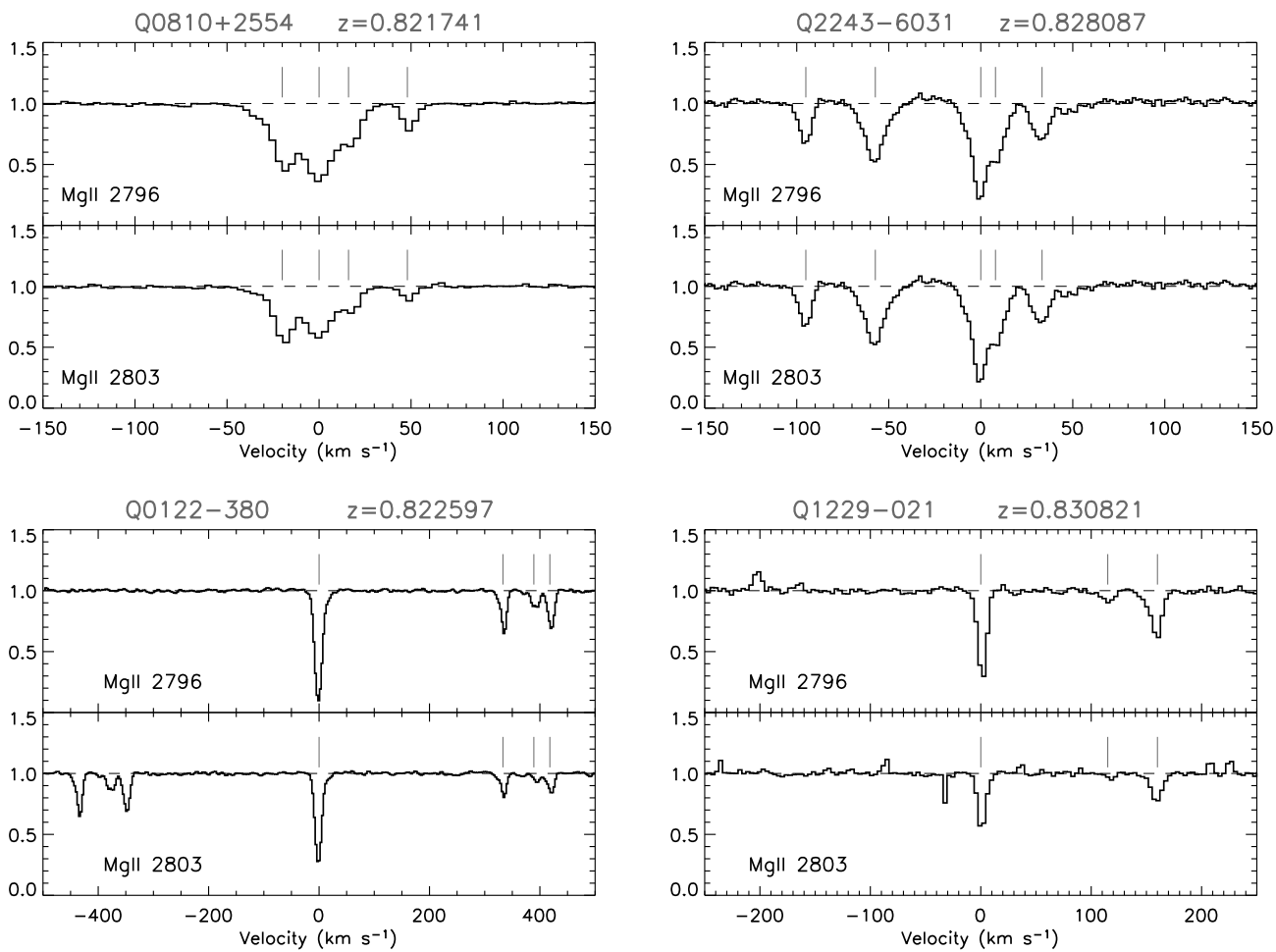}}
\protect
\caption{Absorpiton profiles of the weak systems that were detected in our survey. The top panel in each plot shows the {\MgII}~$\lambda 2796$~{\AA} profile and the bottom panel the {\MgII}~$\lambda 2803$~{\AA} profile. The vertical tick marks represent the center of gaussian fits that were used to determine the equivalent width of the absorbption system. Figures~\ref{fig:2a}-\ref{fig:2e} show only a few examples of the weak {\MgII} systems that we identified in our survey. {\it The absorption profile of all systems identified in our survey will be published in the online version of the journal}}.
\label{fig:2a}
\end{figure*}
\clearpage

\begin{figure*}
\figurenum{2b}
\epsscale{0.8}
\rotatebox{90}{\plotone{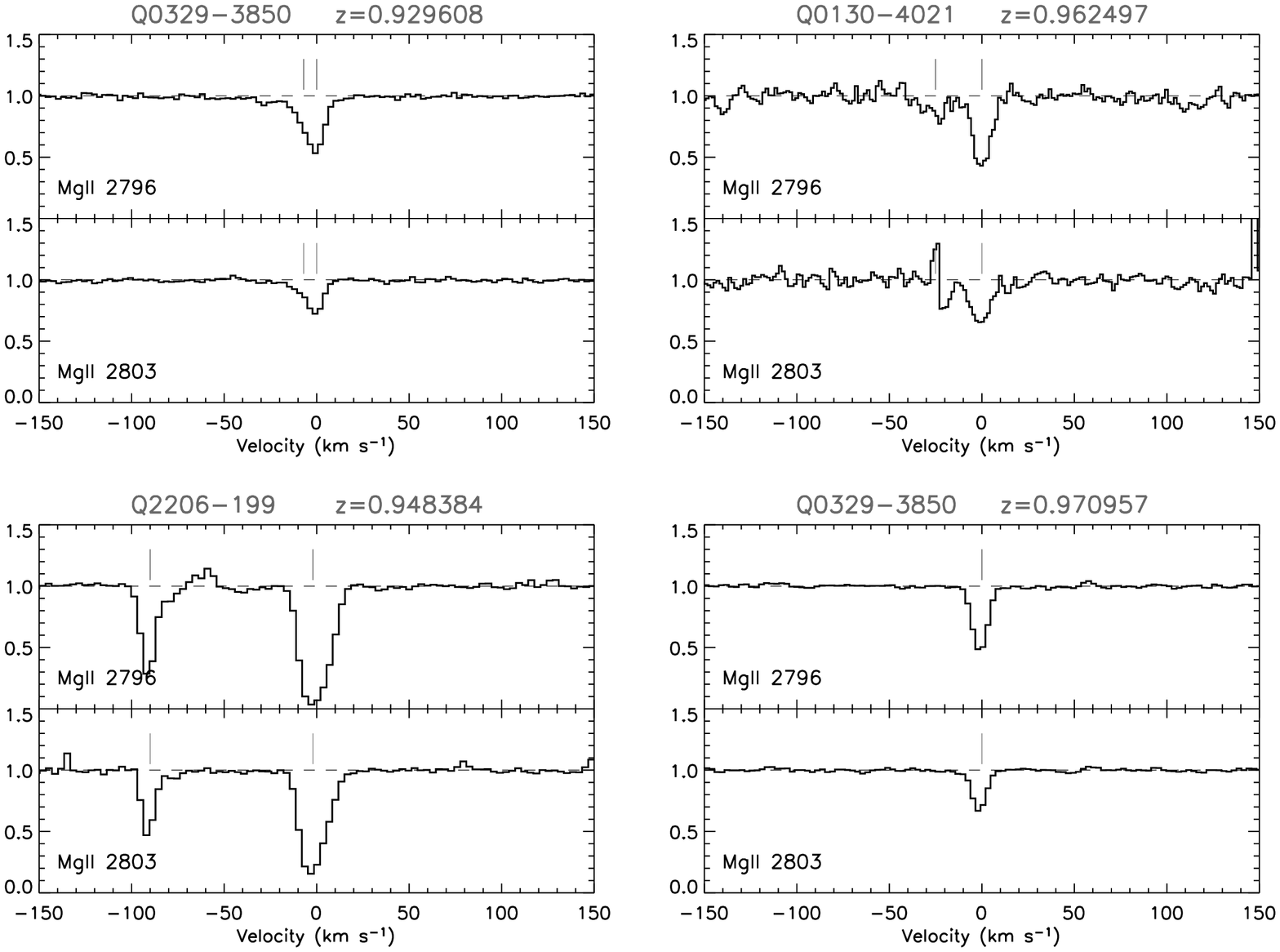}}
\protect
\caption{Contd. Figure~\ref{fig:2a}}
\label{fig:2b}
\end{figure*}
\clearpage

\begin{figure*}
\figurenum{2c}
\epsscale{0.8}
\rotatebox{90}{\plotone{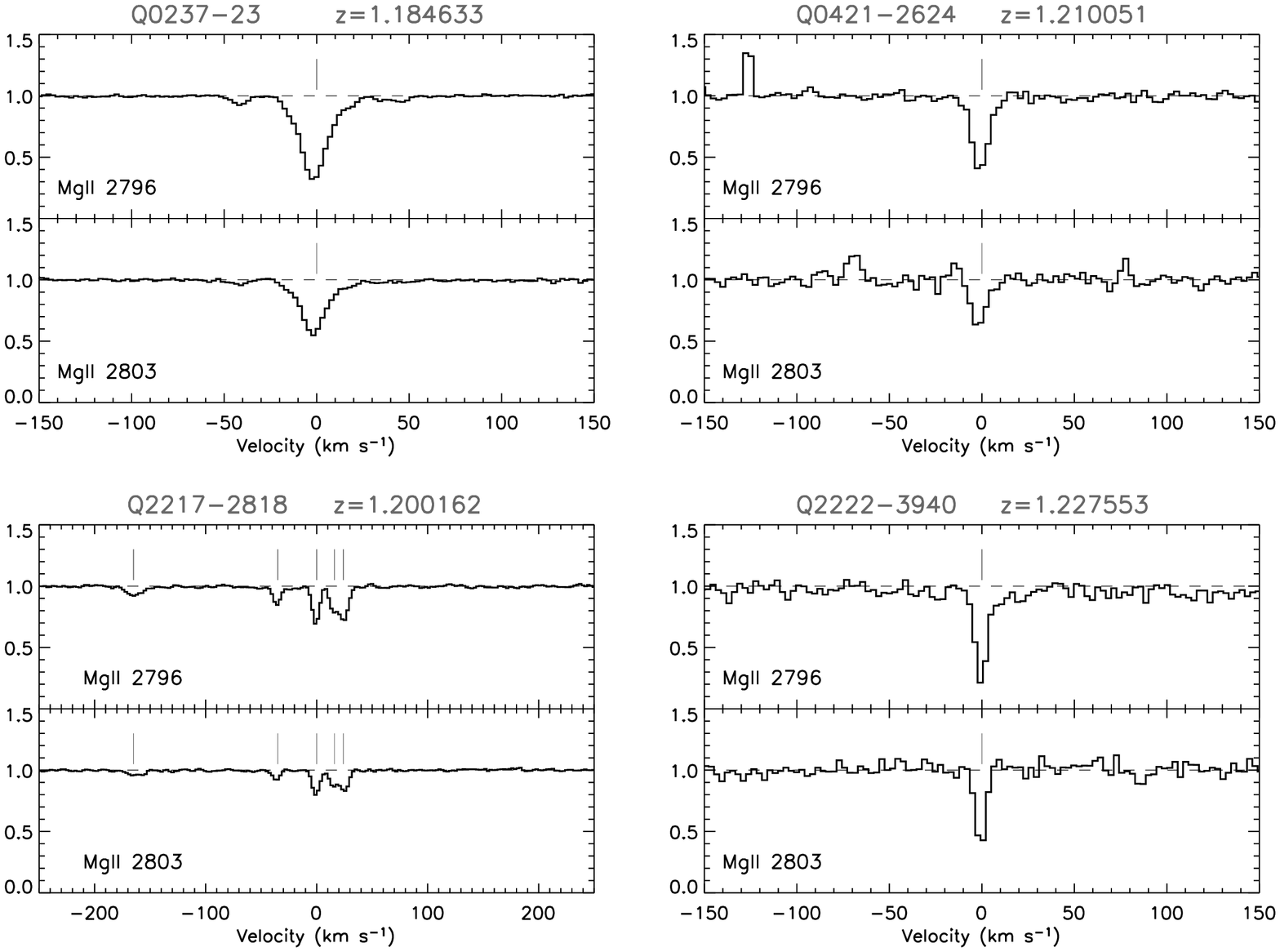}}
\protect
\caption{Contd. Figure~\ref{fig:2a}}
\label{fig:2c}
\end{figure*}
\clearpage

\begin{figure*}
\figurenum{2d}
\epsscale{0.8}
\rotatebox{90}{\plotone{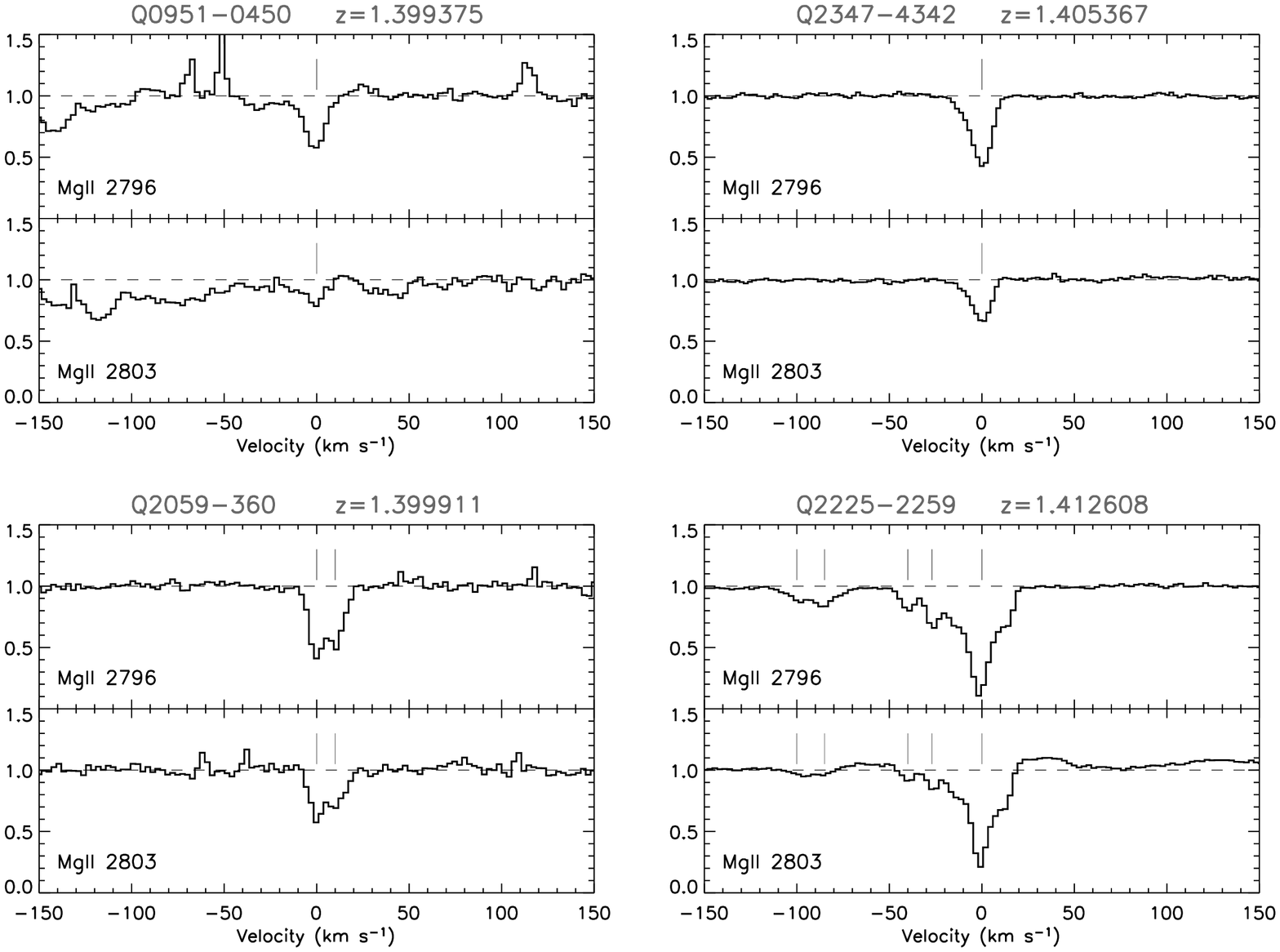}}
\protect
\caption{Contd. Figure~\ref{fig:2a}}
\label{fig:2d}
\end{figure*}
\clearpage

\begin{figure*}
\figurenum{2e}
\epsscale{0.8}
\rotatebox{90}{\plotone{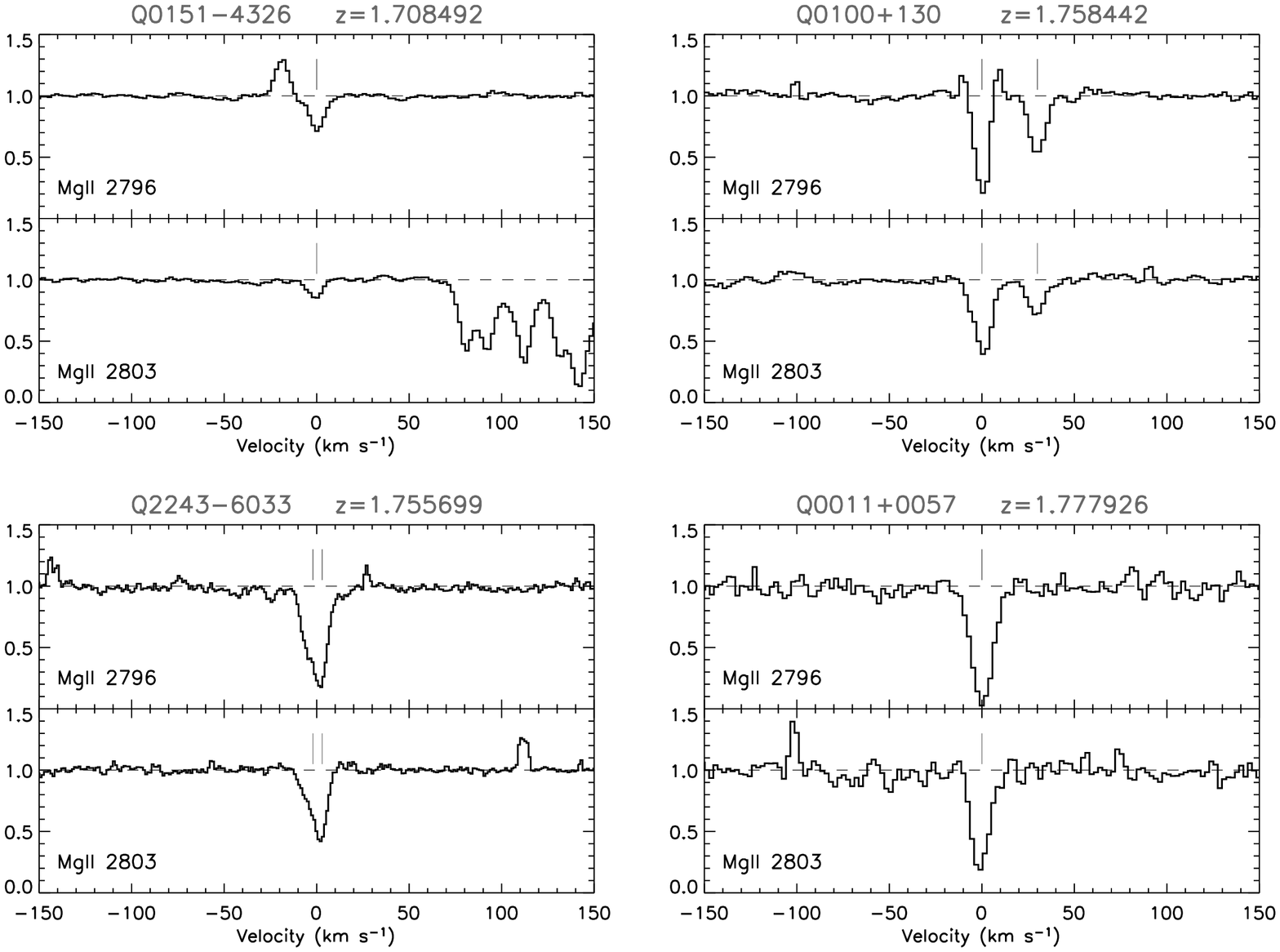}}
\protect
\caption{Contd. Figure~\ref{fig:2a}}
\label{fig:2e}
\end{figure*}
\clearpage

\begin{figure*}
\figurenum{3}
\epsscale{1.0}
\plotone{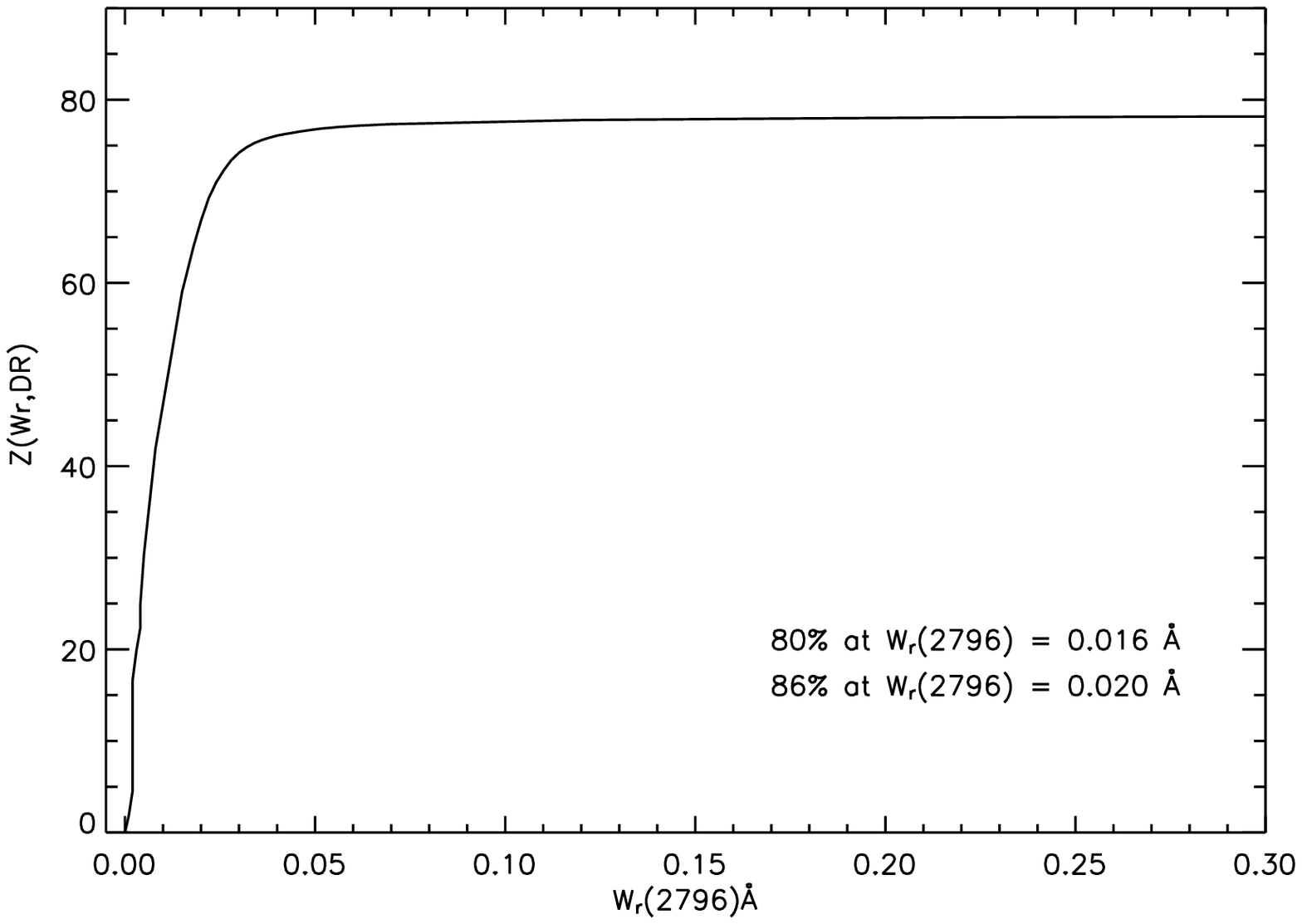}
\protect
\caption{The completeness of the survey is depicted in this figure by plotting the cumulative redshift path as a fucntion of the rest--frame equivalent width, $W_r(2796)$, for the redshift interval $0.4 < z < 2.4$.}
\label{fig:3}
\end{figure*}
\clearpage

\begin{figure*}
\figurenum{4}
\epsscale{1.0}
\plotone{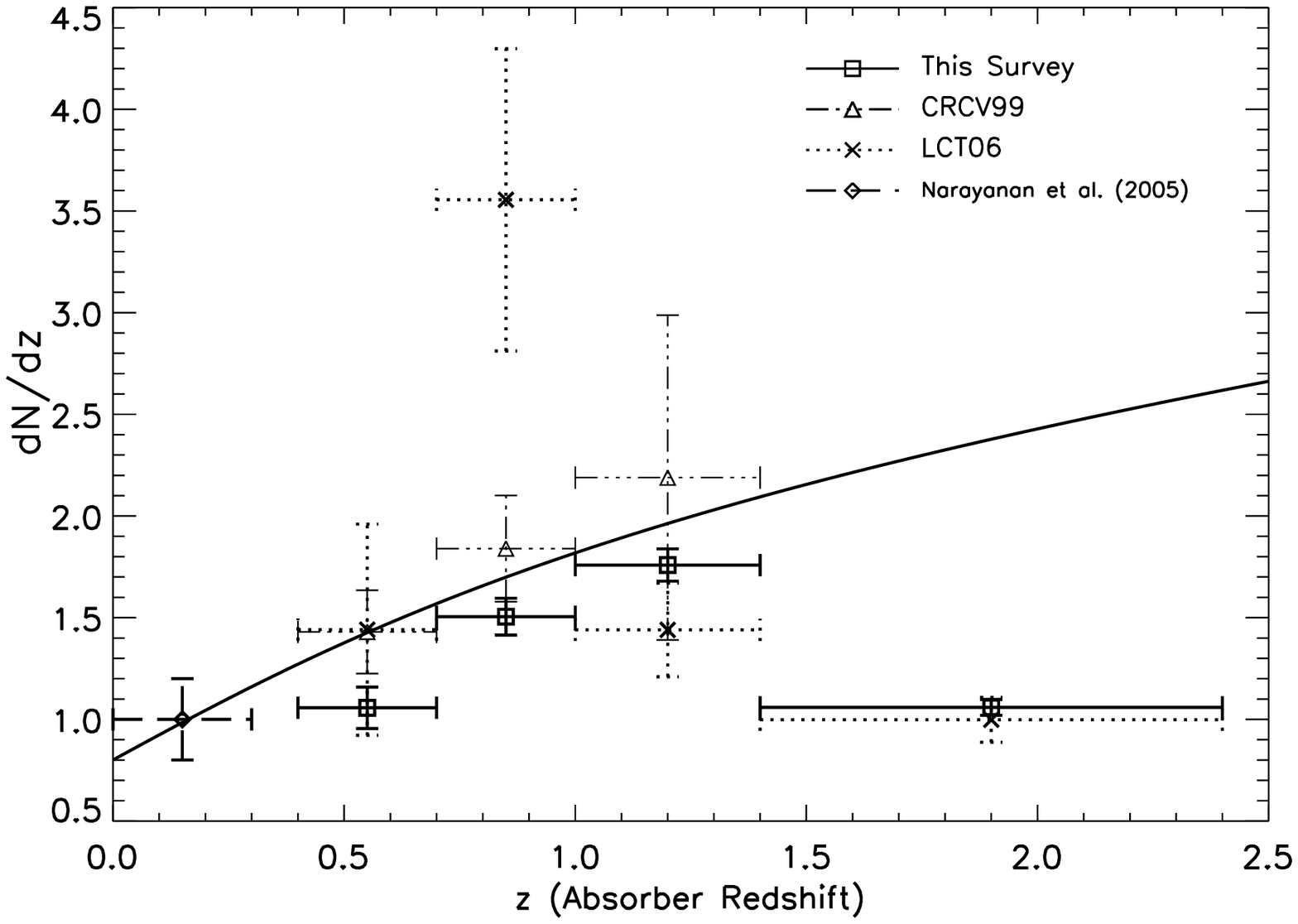}
\protect
\caption{The redshift number density estimates from this survey for the various redshift bins (see Table 3). Also included are the $dN/dz$ constrains from the previous three surveys. \citet{anand05} was a survey for weak {\MgII} systems in the present day universe ($0 < z < 0.3$). The CRCV99 survey covered the redshift window $0.4 < z < 1.4$, and the LCT06 survey covered the same redshift range as this survey ($0.4 < z < 2.4$). The solid curve represents no--evolution expectation in a $\Lambda$CDM universe ($\Omega_m = 0.3$ and $\Omega_{\Lambda} = 0.7$) normalized at $z = 0.9$ and $dN/dz = 1.74$, the normalization used by CRCV99. Systems with $W_r < 0.02$~{\AA} are excluded from our $dN/dz$ estimate, in order to correspond with the equivalent width limits chosen by the CRCV99 and LCT06 surveys.}
\label{fig:4}
\end{figure*}
\clearpage

\begin{figure*}
\figurenum{5}
\epsscale{1.0}
\plotone{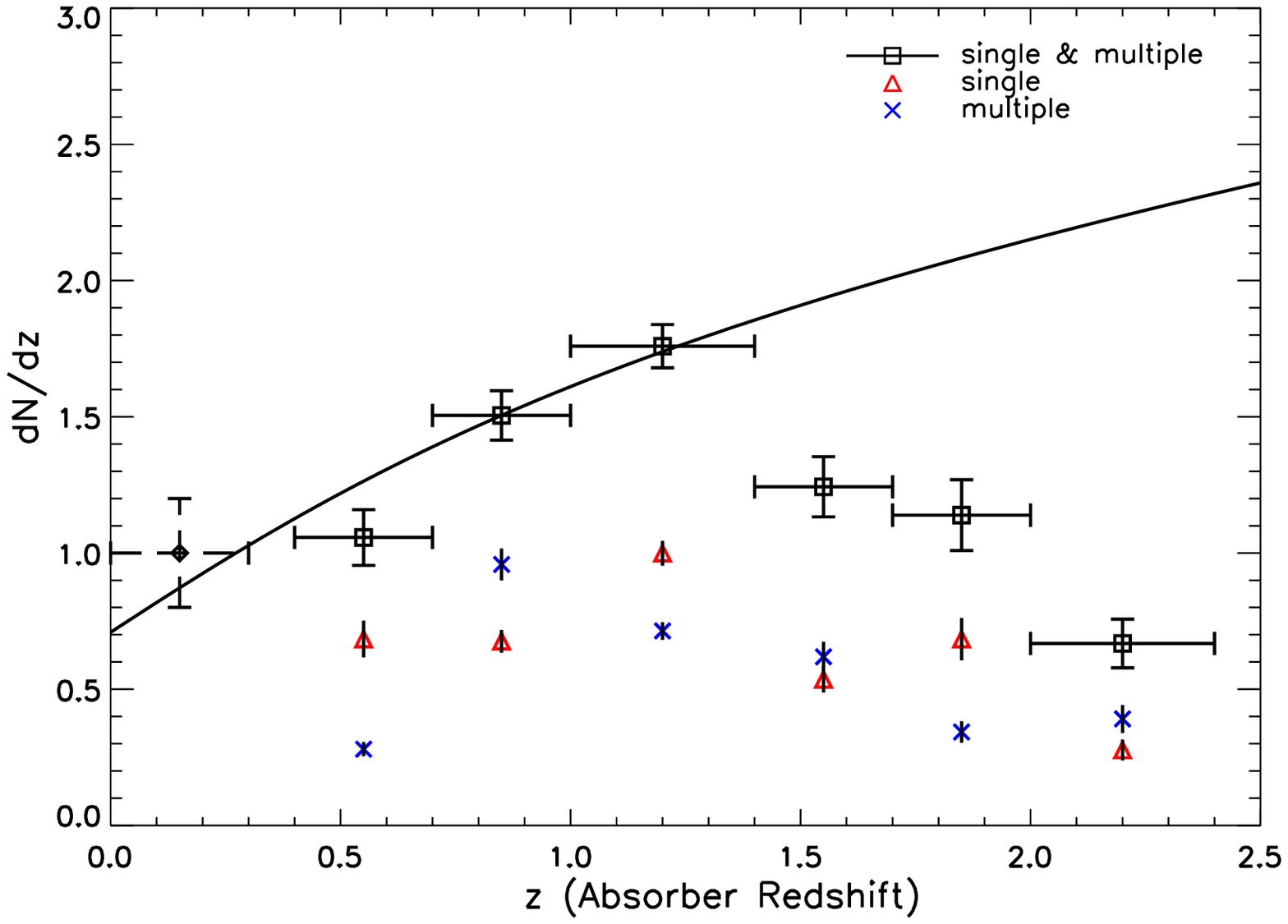}
\protect
\caption{The redshift number density estimates from this survey for the various redshift bins. The $1.4 < z < 2.4$ redshift interval has been split into smaller bin size. The contributions from single and multiple cloud systems are also separately shown. The horizontal bars show the redshift bins for which the $dN/dz$ was calculated. The red triangles and the blue crosses are the $dN/dz$ values for the single and multiple clouds respectively. The $0 < z < 0.3$ data point is from a previous STIS/HST survey by \citet{anand05}. The solid curve shows the expected number density for a non--evolving population of absorbers in a $\Lambda$CDM universe ($\Omega_m = 0.3$ and $\Omega_{\Lambda} = 0.7$) normalized at $z = 0.9$ and $dN/dz = 1.51$.The results distinctly suggest a peak in the number density of weak {\MgII} absorber at $z \sim 1.2$.}
\label{fig:5}
\end{figure*}

\begin{figure*}
\figurenum{6}
\epsscale{0.8}
\rotatebox{90}{\plotone{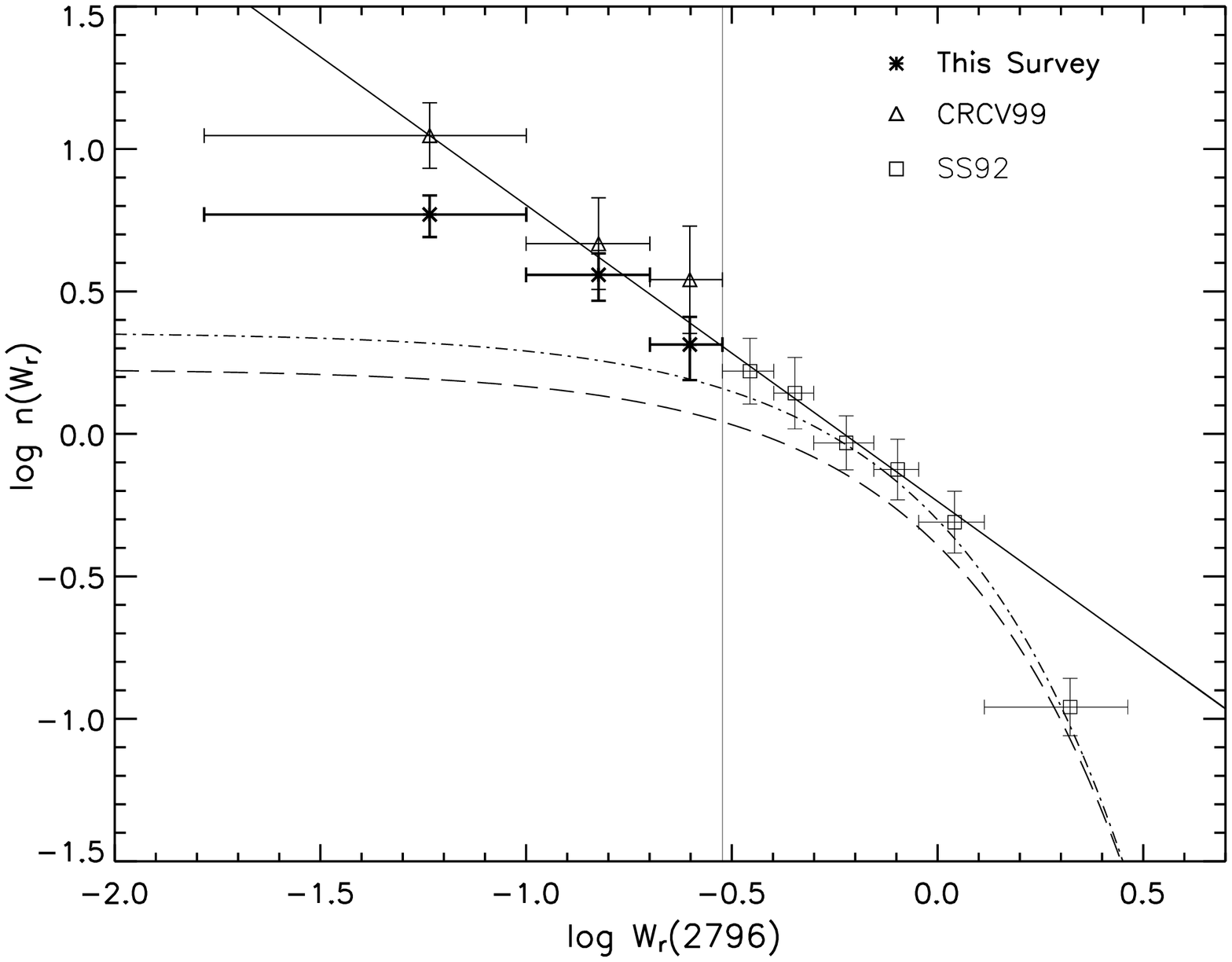}}
\protect
\caption{The equivalent width distribution for strong and weak systems inside the redshift interval $0.4 < z < 1.4$. The three equivalent width bins are [0.0165, 0.1], [0.1, 0.2] and [0.2, 0.3]~{\AA} respectively. The bins were selected to match with the CRCV99 choice of bins. The horizontal bars on each data point represents the equivalent width bins in units of Angstroms. The vertical bars are the error bars for each $n(W_r)$. The points marked with asterisk are results from our survey. The points marked with triangle are from the Keck/HIRES weak {\MgII} survey by CRCV99. The square points represent strong systems from \citet{ss92}. The solid sloped line indicates the power-law function given by equation (4) and based on best-fit parameters derived by CRCV99. The dashed-dot curve indicates the exponential function given by equation (3) and based on best-fit parameters derived by SS92 from fitting strong systems. The dashed curve also represents the exponential function, but with best-fit parameters presented by \citet{nestor05} from their survey of strong systems using the SDSS data. The thin vertical solid line marks the boundary between weak and strong absorbers. Our survey is $78$~\% complete at the lowest equivalent width limit of $W_r=0.0165$~{\AA} in comparision to CRCV99 which is $70$~\% complete at that same limit.}
\label{fig:6}
\end{figure*}

\begin{figure*}
\figurenum{7}
\epsscale{1.0}
\plotone{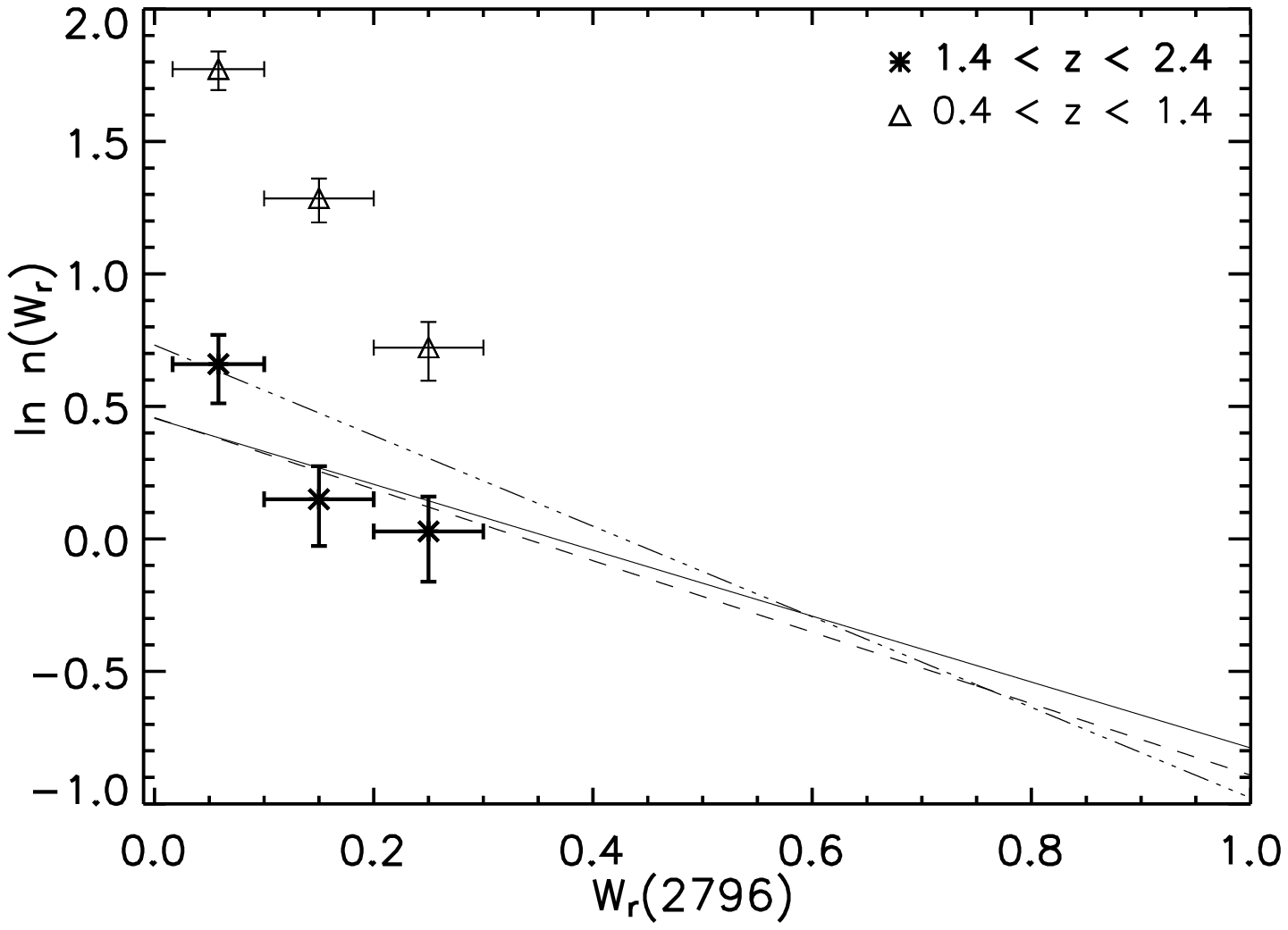}
\protect
\caption{The equivalent width distribution of weak systems within the redshift interval $1.4 < z < 2.4$ from our survey. The horizontal bars represent individual equivalent width bins in Angstroms. The veritcal bars are error bars associated with each $n(W_r)$. The three slanting lines represent the exponential form of equivalent width distribution as described in equation (3) for different redshift intervals with best fit-parameters taken from \citet{nestor05}. The solid line with best-fit parameters of $W^*=0.702$ and $N^*=1.187$ defines the exponential distribution inside the interval $1.311 < z < 2.269$. The dashed line with best-fit parameters of $W^*=0.741$ and $N^*=1.171$ describes the exponential distribution inside the interval $0.871 < z < 1.311$. The dash-dot-dot line with best-fit parameters of $W^*=0.585$ and $N^*=1.216$ is for the redshift interval $0.366 < z < 0.871$. For the sake of comparision, the distribution function $n(W_r)$ for the lower redshift interval $0.4 < z < 1.4$ is also plotted.}
\label{fig:7}
\end{figure*}


\begin{deluxetable}{lcccrrcr}
\tablenum{1}
\tabletypesize{\scriptsize}
\tablecaption{UVES/VLT ARCHIVE QSO DATA SET}
\tablehead{
\colhead{Target} &
\colhead{z$_{QSO}$} &
\colhead{V} &
\colhead{${\lambda}$~{\AA}} &
\colhead{Setting} &
\colhead{t$_{exp}$(s)} &
\colhead{Prog.ID} &
\colhead{PI}
}
\startdata
{TON1480}	 &  $0.614$  &  $17$  &  $3530-6650$  &  $390{\times}564$  &  $18470$  &  {69.A-0371}  &  Savaglio   \\
{Q0300+0048}  &  $0.89$  &  $19.4 (g)$  &  $3070-10000$  &  $437{\times}860$  &  $4500$  &  {267.B-5698}  &  Hutsemekers   \\
  	       &           &  	     &    	         &   $346{\times}580$  &  $4500$  &  {267.B-5698}  &  Hutsemekers	\\
{3c336}  	 &  $0.927$  &  $17.5$ &  $3530-6650$  &  $390{\times}564$  &  $9800$  &  {69.A-0371}  &  Savaglio   \\
{Q0827+243}  &  $0.939$  &  $17.3$  &  $3050-6650$  &  $346{\times}564$  &  $14400$  &  {68.A-0170}  &  Mallen-Ornelas  	\\
  	       &  		 &  		&    	          & $346{\times}564$  &  $19670$  &  {69.A-0371}  &  Savaglio   \\
{Q1229-021}  &  $1.038$  &  $ $  &  $3530-6650$  &  $390{\times}564$  &  $10800$  &  {68.A-0170}  &  Mallen-Ornelas  	\\
{Q1127-14} &  $1.187$  &  $ $  &  $3050-6800$  &  $346{\times}580$  &  $15300$  &  {67.A-0567}  &  Lane   \\
  	       &  		 &  	    &    		  &  $390{\times}564$  &  $9600$  &  {69.A-0371}  &  Savaglio   \\
{Q1243-072}  & $1.286$   & $18.0$ &  $3050-6800$  & $346{\times}580$  & $12000$  & {69.A-0410} & Athreya \\
{Q1453+0029}  &  $1.297$  &  $21.6(g)$  &  $4940-10000$  &  $580$  &  $9000$  &  {267.B-5698}  &  Hutsemekers   \\
  	       &  		 &        &    	 	  &  $860$  &  $9000$  &  {267.B-5698}  &  Hutsemekers   \\
{Q0952+179}  & $1.472$ & $17.2$ & $3050-6650$ & $346{\times}564$  &  $17100$  & 69.A-0371 &  Savaglio \\
{Q2215-0045}  &  $1.475$  &    & $3060-9950$ &  $437{\times}860$  &  $10800$  &  {267.B-5698}  &  Hutsemekers   \\
  	       &  		&  	    &    	    &  $346{\times}580$  &  $10800$  &  {267.B-5698}  &  Hutsemekers   \\
{Q0810+2554}  &  $1.5$  &  $15.4$  &  $3050-6640$  &  $346{\times}564$  &  $48900$  &  {68.A-0107}  &  Reimers   \\
{Q0926-0201}  &  $1.661$  &  $16.4$  &  $3060-10000$  &  $437{\times}860$  &  $3065$  &  {72.A-0446}  &  Murphy   \\
  	         &        &  	    &    	    &   $346{\times}580$  &  $12260$  &  {72.A-0446}  &  Murphy   \\
{Q1629+120}  &  $1.795$  &  $ $  &  $3050-6800$  &  $346{\times}580$  &  $12000$  &  {69.A-0410}  &  Athreya   \\
{Q0141-3932}  &  $1.807$  &  $ $  &  $3060-10000$  &  $437{\times}860$  &  $14400$  &  {67.A-0280}  &  Lopez   \\
  	        &  		  &  	     &    		  &    $346{\times}580$  &  $25200$  &  {67.A-0280(A)}  &  Lopez   \\
{Q0328-272}  &  $1.816$  &  $ $  &  $3500-6630$  &  $390{\times}564$  &  $13200$  &  {072.B-0218}  &  Baker  	\\
{Q2225-2258} & $1.891$ & $17.6$ & $3050-10000$  &  $346{\times}580$  &  $28800$  & {67.A-0280} & Lopez  \\
  	       &  	     &        &   		&  $437{\times}860$  &  $14400$  & {67.A-0280} & Lopez   \\
{Q0136-231}  &  $1.893$  &  $18.8$  &  $3500-6640$  &  $390{\times}564$  &  $9000$  &  {072.B-0218}  &  Baker   \\
  	       &           &  		&    		  &   $520$  &  $3600$  &  {072.B-0218}  &  Baker   \\
{Q0128-2150} & $1.900$ & $15.6$ &  $3050-6800$  & $346{\times}580$  &  $6130$ & {72.A-0446} & Murphy  \\
{Q2044-168}  &  $1.932$  &  $17.36$  &  $3520-9900$  &  $410{\times}800$  &  $6600$  &  {71.B-0106}  &  Pettini  	\\
{Q0429-4901} & $1.940$ & $16.2$ &  $3050-10080$ & $437{\times}860$  &  $18835$  & {66.A-0221} & Lopez \\
  	       &  	     &  	  &  			& $346{\times}580$  & $10800$  & {66.A-0221} & Lopez   \\
{Q0105+061}   & $1.96$   & $17.2$ & $3516-9860$   & $410{\times}800$  &  $9900$ & {71.B-0106} & Pettini \\  
{Q1157+014}  & $1.9997$ & $17.0$ &  $3520-7400$  & $380{\times}580$  & $3600$   & {67.A-0078} & Ledoux \\
  	       &  	     &  	  &  		      & $380{\times}750$  & $7200$  & {67.A-0078} & Ledoux   \\
  	       &  	     &        &   		& $390{\times}580$  & $1800$   & {68.A-0461} & Kanekar \\
{Q1331+170}  &  $2.084$  &  $ $  &  $3050-10000$  &  $346{\times}860$  &  $13500$  &  {67.A-0022}  &  D'Odorico  	\\
{Q0013-0029} &  $2.087$  &  $17$  & $3060-9890$  & $437{\times}750$  &  $10800$  &  {66.A-0624}  &  Ledoux   \\
  	       &  	     &        &    		& $346{\times}580$  &  $19800$  &  {66.A-0624}  &  Ledoux   \\
  	       &  	     &  	  &    		& $346{\times}564$  &  $54000$  &  {267.A-5714}  &  Petitjean  	\\
  	       &         &        &    		& $564$  &  $10800$  &  {267.A-5714}  &  Petitjean  	\\
{Q1246-0217} & $2.106$ & $18.1$ &  $3525-6650$   & $390{\times}564$  & $5400$   & {67.A-0146} & Vladilo \\
{Q1341-1020}  &  $2.135$  &  $17.1$  & $3060-10400$ &  $346{\times}580$  &  $32400$  &  {160.A-0106}  &  Bergeron  	\\
{Q0010-0012}  &  $2.145$  &  $19.43$  &  $3050 - 6650$  &  $346{\times}564$  &  $5400$  &  {68.A-0600}  &  Ledoux  	\\
{Q2222-3939}  &  $2.18$  &  $17.9$  &  $3530-6640$  &  $390{\times}564$  &  $1800$  &  {072.A-0442}  &  Lopez   \\
{Q0122-380}  &  $2.200$  &  $17.1$  &  $3060-10190$  &  $346{\times}580$  &  $21600$  &  {160.A-0106}  &  Bergeron  	\\
{Q1444+014}  &  $2.206$  &  $ $  &  $3520-5830$  &  $390{\times}564$  &  $18000$  &  {65.O-0158}  &  Pettini  	\\
  	       &  		 &  	    &    		  &  $380{\times}564$  &  $10800$  &  {67.A-0078}  &  Ledoux   \\
  	       &           &  	    &    		  &  $390{\times}564$  &  $10800$  &  {69.B-0108}  &  Srianand  	\\
  	       &           &  	    &    		  &  $390{\times}564$  &  $14400$  &  {71.B-0136}  &  Srianand  	\\
{Q1448-232} &  $2.215$  &  $17$  & $3060-10070$ &  $346{\times}580$  &  $28800$  &  {160.A-0106}  &  Bergeron  	\\
  	       &  		 &  	    &    		  &  $437{\times}860$  &  $21600$  &  {160.A-0106} & Bergeron \\
{Q0237-23}  &  $2.223$  &  $16.8$  & $3060-10070$ &  $346{\times}580$  &  $21600$  &  {160.A-0106}  &  Bergeron  	\\
  	       &  		 &  	    &    		  &  $437{\times}860$  &  $21600$  & {160.A-0106} & Bergeron \\
{Q0549-213}  &  $2.245$  &  $20$  &  $3500-6640$  &  $390{\times}564$  &  $12000$  &  {072.B-0218}  &  Baker  	\\
{Q0425-5214}  &  $2.25$  &  $17.8$  &  $3520-6645$  &  $390{\times}564$  &  $1800$  &  {072.A-0442}  &  Lopez   \\
{Q0049-2820}  &  $2.256$  &  $18.42$  &  $3520-6645$  &  $390{\times}580$  &  $1800$  &  {072.A-0442}  &  Lopez  	\\
{Q0421-2624}  &  $2.277$  &  $18.08$  &  $3520-6640$  &  $390{\times}564$  &  $1800$  &  {072.A-0442}  &  Lopez   \\
{Q0001-2340}  &  $2.28$  &  $16.7$  &  $3060-10070$ &  $346{\times}580$  &  $21600$  &  {160.A-0106}  &  Bergeron  	\\
  	       &  		 &  		  &    	  &  $437{\times}860$  &  $21600$  & {160.A-0106}  & Bergeron \\
{Q1114-220}  &  $2.282$  &  $20.2$  &  $3540-6645$  &  $390{\times}564$  &  $14120$  &  {71.B-0081}  &  Baker   \\
{Q0011+0055}  &  $2.31$  &  $19.1(g)$  &  $3770-10000$  &  $437{\times}860$  &  $7200$  &  {267.B-5698}  &  Hutsemekers   \\
{Q0551-3637}  &  $2.318$  &  $17.0$  &  $3060-9370$  &  $437{\times}750 $  &  $8100$  &  {66.A-0624}  &  Ledoux   \\
  	       &  		  &  	     &    		  &   $346{\times}580$  &  $18000$  &  {66.A-0624}  &  Ledoux   \\
{Q2116-358}  &  $2.341$  &  $17$  &  $3530-6640$  &  $390{\times}564$  &  $7200$  &  {65.O-0158}  &  Pettini  	\\
{Q0042-2930}  &  $2.388$  &  $17.81$  &  $3530-6800$  &  $390{\times}580$  &  $1800$  &  {072.A-0442}  &  Lopez  	\\
{Q0109-3518}  &  $2.405$  &  $16.6$  & $3060-10070$  &  $346{\times}580$  &  $25200$  &  {160.A-0106}  &  Bergeron  	\\
  	        &           &  		 &    	  &  $437{\times}860$  &  $21600$  &  {160.A-0106} & Bergeron 	\\
{Q1122-1648} &  $2.405$  &  $17.7$  & $3060-10070$ &  $346{\times}580$  &  $26400$  &  {Sci. Veri}  &    	\\
  	        &    	  &          &    	  &  $437{\times}860$  &  $27000$  &  {Sci. Veri}  &   	\\
{Q2217-2818} &  $2.406$  &  $16.0$    & $3060-9890$ &  $346{\times}580$  &  $16200$  &  {Comm.}  &    	\\
  	        &  		  &  		 &    	  &  $390{\times}564$  &  $10800$  & {Comm.} &     	\\
{Q2132-433}  &  $2.420$  &  $18.18$  &  $3500-6640$  &  $390{\times}564$  &  $3600$  &  {65.O-0158}  &  Pettini  	\\
{Q0329-385}  &  $2.435$   &  $17.2$  &  $3070-8500$  &  $346{\times}580$  &  $21600$  &  {160.A-0106}  &  Bergeron  	\\
  	       &    	  &	       &    	  &  $437{\times}860$  &  $21600$  &  {160.A-0106}  & Bergeron   	\\
{Q2314-409}  &  $2.448$  &  $17.9$  &  $3520-6640$  &  $390{\times}564$  &  $13680$  &  {267.A-5707}  &  Ellison  	\\
{Q1158-1843}  &  $2.448$  &  $16.9$  & $3070-10070$ &  $346{\times}580$  &  $21600$  &  {160.A-0106}  &  Bergeron  	\\
  	       &            &  		 &    	  &  $437{\times}860$  &  $21600$  &  {160.A-0106}  & Bergeron    	\\
{Q2206-199}  &  $2.56$  &  $17.3$  &  $3420-6640$  &  $390{\times}564$  &  $17100$  &  {65.O-0158}  &  Pettini   \\
{Q1140+2711} & $2.630$ & $17.0$ &  $3775-10000$   &  $437{\times}860$  &  $45400$  & 69.A-0246 & Reimers \\
{Q0453-423}  &  $2.657$  &  $17.3$  &    	  &  $346{\times}580$  &  $28800$  &  {160.A-0106}  &  Bergeron  	\\
  	       &        &           &    	  &  $437{\times}860$  &  $28800$  &  {160.A-0106}  & Bergeron 	\\
{Q0100+1300}  &  $2.681$  &  $16.57$  &  $3520-10000$  &  $390{\times}860$  &  $7200$  &  {67.A-0022}  &  D'Odorico  	\\
{Q0329-255}  &  $2.685$  &  $17.51$  & $3060-10070$  &  $346{\times}580$  &  $46800$  &  {160.A-0106}  &  Bergeron  	\\
  	       &  	&  	&    	  	  &  $437{\times}860$  &  $39600$  &  {160.A-0106}  &  Bergeron 	\\
{Q0151-4326}  &  $2.74$  &  $17.19$  &  $3060-10070$  &  $346{\times}580$  &  $28800$  &  {160.A-0106}  &  Bergeron  	\\
  	       &  		&  		  &    	  &  $437{\times}860$  &  $32400$  &  {160.A-0106}  & Bergeron 	\\
{Q0002-422}  &  $2.76$  &  $17.2$  & $3160-10070$   &  $346{\times}580$  &  $28800$  &  {160.A-0106}  &  Bergeron  	\\
  	       &  	&  	    &    	  &  $437{\times}860$  &  $39600$ &  {160.A-0106}   & Bergeron  \\
{Q1151+068}  & $2.762$ & $18.6$ &  $3705-10000$ & $346{\times}580$  &  $10800$  & {65.O-0158} & Pettini \\
  	       &          &  		  &    	  & $437{\times}860$ & $10800$  & {65.O-0158} & Pettini  \\
{Q0112+0300}  &  $2.81$  &  $ $  &  $3540-6800$  &  $390{\times}580$  &  $3600$  &  {66.A-0624}  &  Ledoux   \\
{Q2347-4342}  &  $2.88$  &  $16.3$  & $3100-10070$  &  $346{\times}580$  &  $21600$  &  {160.A-0106}  &  Bergeron  	\\
  	       &  		  &  		 &    	  &  $437{\times}860$  &  $28800$  &  {160.A-0106} & Bergeron \\
{Q1337+113}  &  $2.919$  &  $18.7$  &  $3540-9380$  &  $410{\times}750$  &  $10800$  &  {67.A-0078}  &  Ledoux   \\
{Q2243-6031}  &  $3.01$  &  $18.3$  &  $3140-10000$  &  $346{\times}580$  &  $14400$  &  {65.O-0411}  &  Lopez   \\
  	        &  		&  		&    		  &    $437{\times}860$  &  $11400$  &  {65.O-0411}  &  Lopez   \\
{Q0130-4021}  &  $3.023$  &  $17.02$  &  $3550-6800$  &  $390{\times}580$  &  $3065$  &  {70.B-0522}  &  Bomans  	\\
{Q0102-1902}  &  $3.04$  &  $ $  &  $3620-10000$  &  $390{\times}564$  &  $3600$  &  {67.A-0146}  &  Vladilo   \\
  	       &  		 &  	    &    		  &   $437{\times}860$  &  $10800$  &  {67.A-0146}  &  Vladilo   \\
{Q0940-1050}  &  $3.083$  &  $16.6$  & $3110-10070$  &  $346{\times}580$  &  $18000$  &  {160.A-0106}  &  Bergeron  	\\
  	       &            &      &    		  &  $437{\times}860$  &  $14400$  & {160.A-0106} & Bergeron  \\
{Q2059-360}  &  $3.092$  &  $18.62$  &  $3750-9280$  &  $433{\times}740$  &  $18000$  &  {67.A-0078}  &  Ledoux   \\
{Q0058-2914}  &  $3.093$  &  $18.7$  &  $3550-10000$  &  $390{\times}580$  &  $5400$  &  {66.A-0624}  &  Ledoux   \\
  	       &      	 & 	    &    		  &   $437{\times}860$  &  $34200$  &  {67.A-0146}  &  Vladilo   \\
{Q0420-388}  &  $3.117$  &  $16.9$  & $3760-10070$ &  $390{\times}564$  &  $28800$  &  {160.A-0106}  &  Bergeron  	\\
  	       &    	 &          &    		  &  $437{\times}860$  &  $28800$  &  {160.A-0106}  & Bergeron 	\\
{Q2204-408}  &  $3.155$  &  $17.57$  &  $3520-6800$  &  $390{\times}580$  &  $6600$  &  {71.B-0106}  &  Pettini   \\
  	       &  	       &          &  	           &  $580$       &  $9900$  &   {71.B-0106} &  Pettini   \\
{Q2126-158} & $3.28$   & $17.3$  &  $3520-9600$  &  $390{\times}564$  &  $32400$ &  {160.A-0106} & Bergeron \\
	      & 	 &	   & 		     &  $437{\times}860$  &  $28800$ &  {160.A-0106} & Bergeron \\
{Q1209+0919}  & $3.3$  &  $18.5$  &  $3520-7770$  & $390{\times}564$  &  $5400$  & {67.A-0146} & Vladilo \\
	      &        &          &               & $437{\times}860$  &  $14765$ & {67.A-0146} & Vladilo \\
	      &        &          &               & $390{\times}580$  &  $3600$ & {73.B-0787} & Dessauges-Zavadsky \\
{CTQ0298}     & $3.37$ &  $17.60$ &  $3520-8550$  & $390{\times}564$  &  $10800$ & {68.A-0492} & D'Odorico \\
	      &	       &          & 		  & $436{\times}800$  &  $14400$ & {68.A-0492} & D'Odorico \\
{Q0055-269}   & $3.66$ &  $17.47$ & $3060-7505$   & $346{\times}580$  &  $17000$ & {65.O-0296} & D'Odorico \\
	      &        &          &               & $437{\times}800$  &  $16300$ & {65.O-0296} & D'Odorico \\ 
	      &        &          &               & $346{\times}565$  &  $9300$ & {65.O-0296} & D'Odorico \\
	      &        &          &               & $565$  &  $8800$ & {65.O-0296} & D'Odorico \\
{Q1418-064}& $3.689$ & $18.5$    & $3765-9945$   & $437{\times}860$  &  $3575$ & {69.A-0051} & Pettini \\  
	      &        &          &               & $437{\times}860$  &  $3814$ & {71.A-0539} & Kanekar \\ 
	      &        &          &               & $437{\times}860$  &  $7150$ & {71.A-0067} & Ellison \\  
{Q1621-0042}  &  $3.7$  &  $ $  &  $3530-6800$  &  $390{\times}580$  &  $28105$  &  {075.A-0464}  &  Kim   \\
{Q2000-330} & $3.773$ & $17.3$  & $3495-9945$   & $580$  &  $3600$ & {65.O-0299} & D'Odorico \\ 
	      &        &          &               & $390{\times}560$  &  $32400$ & {166.A-0106} & Bergeron \\ 
	      &        &          &               & $437{\times}860$  &  $32400$ & {166.A-0106} & Bergeron \\ 
{Q1108-0747}& $3.922$ & $18.10$ & $3765-9945$   & $520$  &  $4500$ & {67.A-0022} & D'Odorico \\
	      &        &          &               & $520$  &  $6000$ & {68.A-0492} & D'Odorico \\
	      &	       &	  &		  & $480{\times}800$ & $7200$ & {68.A-0492} & D'Odorico \\
	      &        &          &               & $437$  &  $3598$ & {68.B-0115} & Molaro \\
	      &        &          &               & $580$  &  $4800$ & {68.B-0115} & Molaro \\
	      &        &          &               & $437{\times}860$  &  $9600$ & {68.B-0115} & Molaro \\
{Q0401-1711} & $4.23$ & $18.7$   & $4785-9880$   & $580$  &  $6000$ & {074.A-0306} & D'Odorico \\
	      &        &          &               & $860$  &  $6625$ & {074.A-0306} & D'Odorico \\
	      &        &          &               & $800$  &  $2617$ & {71.B-0106} & Pettini \\
{Q2344+0342} & $4.239$ &  $18.6$ & $4635-9890$   & $800$  &  $9000$ & {65.O-0296} & D'Odorico \\
	      &        &          &               & $565$  &  $4500$ & {65.O-0296} & D'Odorico \\
{Q0951-0450}  & $4.369$ & $18.9$  & $4785-6805$   & $580$  &  $30960$ & {072.A-0558} & Vladilo \\
{Q1114-0822}  & $4.495$ & $19.4$  & $4785-10000$  & $580{\times}860$  &  $14415$ & {074.A-0801} & Molaro \\
{Q1202-0725}  & $4.694$ & $17.5$  & $3535-10000$  & $437{\times}860$  &  $10958$ & {66.A-0594} & Molaro \\
	      &        &          &               & $390{\times}580$  &  $38693$ & {66.A-0594} & Molaro \\
	      &        &          &               & $580$  &  $18000$ & {166.A-0106} & Bergeron \\
	      &        &          &               & $860$  &  $7200$ & {166.A-0106} & Bergeron \\
	      &        &          &               & $860$  &  $17000$ & {71.B-0106} & Pettini \\
\enddata
\label{tab:1}
\tablecomments{This table provides details of the archived VLT/UVES spectra that were used in our survey. The second and third column lists the redshift of the quasar and its magnitude as given by Simbad and/or NED data base. The fourth column gives the wavelength coverage for each case. The fifth column is the cross-disperser settings that were used for the various exposures. The sixth column gives the total exposure time (in seconds). The program ID and the PI of the program are listed in the final two columns.}
\end{deluxetable}

\begin{deluxetable}{lccccc}
\tablenum{2}
\tablewidth{0pt}
\tablecaption{WEAK {\MgII} SYSTEMS DETECTED}
\tablehead{
\colhead{QSO} &
\colhead{z$_{abs}$} &
\colhead{W$_r(2796)$} &
\colhead{W$_r(2803)$} &
\colhead{DR} &
\colhead{Z($W_r$,DR)} 
}
\startdata
{Q1127-145} &     $0.190587$   &     $0.137 ~{\pm}~ 0.009$   &    $0.054 ~{\pm}~ 0.007$   &    $2.54 ~{\pm}~ 0.37$   &   $-$	\\
{Q0827+243} &     $0.259000$   &     $0.273 ~{\pm}~ 0.006$   &    $0.201 ~{\pm}~ 0.005$   &    $1.36 ~{\pm}~ 0.05$   &   $-$	\\
{Q1127-145} &     $0.328258$   &     $0.028 ~{\pm}~ 0.004$   &    $0.018 ~{\pm}~ 0.000$   &    $1.56 ~{\pm}~ 0.22$   &   $-$	\\
{Q0141-3932} &     $0.340005$   &     $0.227 ~{\pm}~ 0.003$   &    $0.116 ~{\pm}~ 0.003$   &    $1.96 ~{\pm}~ 0.06$   &  $-$ 	\\
{Q1444+014} &     $0.444019$   &     $0.228 ~{\pm}~ 0.003$   &    $0.113 ~{\pm}~ 0.003$   &    $2.02 ~{\pm}~ 0.06$   &   77.66	\\
{Q0001-2340}  &     $0.452414$   &     $0.105 ~{\pm}~ 0.001$   &    $0.071 ~{\pm}~ 0.001$   &    $1.48 ~{\pm}~ 0.03$   &   77.33	\\
{Q0011+0055} &     $0.487243$   &     $0.244 ~{\pm}~ 0.019$   &    $0.129 ~{\pm}~ 0.016$   &    $1.89 ~{\pm}~ 0.28$   &   77.70	\\
{Q0551-3637} &     $0.505268$   &     $0.083 ~{\pm}~ 0.007$   &    $0.062 ~{\pm}~ 0.022$   &    $1.34 ~{\pm}~ 0.49$   &   77.15	\\
{Q1158-1843} &     $0.506041$   &     $0.022 ~{\pm}~ 0.001$   &    $0.013 ~{\pm}~ 0.001$   &    $1.69 ~{\pm}~ 0.15$   &   68.48	\\
{Q1444+014} &     $0.509653$   &     $0.140 ~{\pm}~ 0.005$   &    $0.083 ~{\pm}~ 0.005$   &    $1.69 ~{\pm}~ 0.12$   &   77.41	\\
{Q2116-358} &     $0.539154$   &     $0.102 ~{\pm}~ 0.013$   &    $0.084 ~{\pm}~ 0.004$   &    $1.21 ~{\pm}~ 0.17$   &   77.35	\\
{Q0328-272} &     $0.570827$   &     $0.168 ~{\pm}~ 0.008$   &    $0.098 ~{\pm}~ 0.007$   &    $1.71 ~{\pm}~ 0.15$   &   77.56	\\
{Q0429-4901} &     $0.584249$   &     $0.017 ~{\pm}~ 0.002$   &    $0.008 ~{\pm}~ 0.001$   &    $2.13 ~{\pm}~ 0.36$   &   59.34	\\
{Q2217-2818} &     $0.599512$   &     $0.114 ~{\pm}~ 0.001$   &    $0.067 ~{\pm}~ 0.003$   &    $1.70 ~{\pm}~ 0.08$   &   77.34	\\
{Q0013-0029} &     $0.635069$   &     $0.162 ~{\pm}~ 0.022$   &    $0.091 ~{\pm}~ 0.009$   &    $1.78 ~{\pm}~ 0.30$   &   77.5	\\
{Q0001-2340} &     $0.685957$   &     $0.033 ~{\pm}~ 0.001$   &    $0.018 ~{\pm}~ 0.001$   &    $1.83 ~{\pm}~ 0.12$   &   74.19	\\
{Q1229-021} &     $0.700377$   &     $0.010 ~{\pm}~ 0.001$   &    $0.008 ~{\pm}~ 0.002$   &    $1.25 ~{\pm}~ 0.34$   &   46.52	\\
{3c336}	 &     $0.702901$   &     $0.028 ~{\pm}~ 0.004$   &    $0.022 ~{\pm}~ 0.003$   &    $1.27 ~{\pm}~ 0.25$   &   72.94	\\
{Q0151-4326} &     $0.737248$   &     $0.022 ~{\pm}~ 0.001$   &    $0.019 ~{\pm}~ 0.001$   &    $1.16 ~{\pm}~ 0.08$   &   68.71	\\
{Q1229-021} &     $0.756921$   &     $0.298 ~{\pm}~ 0.004$   &    $0.238 ~{\pm}~ 0.004$   &    $1.25 ~{\pm}~ 0.03$   &   77.82	\\
{Q1229-021} &     $0.768862$   &     $0.026 ~{\pm}~ 0.002$   &    $0.011 ~{\pm}~ 0.001$   &    $2.36 ~{\pm}~ 0.28$   &   68.28	\\
{Q0109-3518} &     $0.769646$   &     $0.033 ~{\pm}~ 0.001$   &    $0.018 ~{\pm}~ 0.001$   &    $1.83 ~{\pm}~ 0.12$   &   74.19	\\
{Q2116-358} &     $0.775270$   &     $0.238 ~{\pm}~ 0.039$   &    $0.110 ~{\pm}~ 0.014$   &    $2.16 ~{\pm}~ 0.45$   &   77.66	\\
{Q2217-2818} &     $0.786572$   &     $0.207 ~{\pm}~ 0.003$   &    $0.114 ~{\pm}~ 0.001$   &    $1.82 ~{\pm}~ 0.03$   &   77.62	\\
{Q2132-433} &     $0.793570$   &     $0.184 ~{\pm}~ 0.008$   &    $0.118 ~{\pm}~ 0.008$   &    $1.56 ~{\pm}~ 0.13$   &   77.61	\\
{Q0042-2930} &     $0.798665$   &     $0.243 ~{\pm}~ 0.008$   &    $0.149 ~{\pm}~ 0.006$   &    $1.64 ~{\pm}~ 0.09$   &   77.71	\\
{Q1122-1648} &     $0.806215$   &     $0.245 ~{\pm}~ 0.001$   &    $0.154 ~{\pm}~ 0.001$   &    $1.59 ~{\pm}~ 0.01$   &   77.72	\\
{Q1158-1843} &     $0.818146$   &     $0.063 ~{\pm}~ 0.001$   &    $0.038 ~{\pm}~ 0.001$   &    $1.66 ~{\pm}~ 0.05$   &   76.67	\\
{Q0810+2554} &     $0.821741$	&     $0.252 ~{\pm}~ 0.002$	&    $0.166 ~{\pm}~ 0.002$	&    $1.52 ~{\pm}~ 0.02$   &   77.76	\\
{Q0122-380} &     $0.822597$   &     $0.253 ~{\pm}~ 0.007$   &    $0.138 ~{\pm}~ 0.012$   &    $1.83 ~{\pm}~ 0.17$   &   77.73	\\
{Q2243-6031} &     $0.828087$   &     $0.242 ~{\pm}~ 0.003$   &    $0.135 ~{\pm}~ 0.004$   &    $1.79 ~{\pm}~ 0.06$   &   77.70	\\
{Q1229-021} &     $0.830821$   &     $0.126 ~{\pm}~ 0.004$   &    $0.071 ~{\pm}~ 0.003$   &    $1.77 ~{\pm}~ 0.09$   &   77.34	\\
{Q2225-2258} &     $0.831374$   &     $0.031 ~{\pm}~ 0.002$   &    $0.020 ~{\pm}~ 0.002$   &    $1.55 ~{\pm}~ 0.18$   &   73.80	\\
{Q0810+2554} &     $0.831727$	&     $0.171 ~{\pm}~ 0.002$	&    $0.084 ~{\pm}~ 0.002$	&    $2.04 ~{\pm}~ 0.05$	&   77.5	\\
{Q2314-409} &     $0.843114$   &     $0.044 ~{\pm}~ 0.003$   &    $0.028 ~{\pm}~ 0.004$   &    $1.57 ~{\pm}~ 0.25$   &   75.81	\\
{Q0013-0029} &     $0.857469$   &     $0.142 ~{\pm}~ 0.004$   &    $0.123 ~{\pm}~ 0.004$   &    $1.15 ~{\pm}~ 0.05$   &   77.45	\\
{Q0453-4230} &     $0.895865$   &     $0.034 ~{\pm}~ 0.001$   &    $0.019 ~{\pm}~ 0.001$   &    $1.79 ~{\pm}~ 0.11$   &   74.45	\\
{Q0109-3518} &     $0.895295$   &     $0.020 ~{\pm}~ 0.001$   &    $0.011 ~{\pm}~ 0.001$   &    $1.82 ~{\pm}~ 0.19$   &   65.54	\\
{Q0122-380} &     $0.910146$   &     $0.061 ~{\pm}~ 0.004$   &    $0.028 ~{\pm}~ 0.011$   &    $2.18 ~{\pm}~ 0.87$   &   76.42	\\
{Q0102-1902} &     $0.916743$   &     $0.294 ~{\pm}~ 0.010$   &    $0.228 ~{\pm}~ 0.012$   &    $1.30 ~{\pm}~ 0.08$   &   77.82	\\
{Q03290-3850} &     $0.929608$   &     $0.072 ~{\pm}~ 0.007$   &    $0.037 ~{\pm}~ 0.004$   &    $1.95 ~{\pm}~ 0.28$   &   76.73	\\
{Q2206-199} &     $0.948384$   &     $0.255 ~{\pm}~ 0.002$   &    $0.180 ~{\pm}~ 0.002$   &    $1.42 ~{\pm}~ 0.02$   &   77.77	\\
{Q0130-4021} &     $0.962497$   &     $0.089 ~{\pm}~ 0.004$   &    $0.060 ~{\pm}~ 0.004$   &    $1.48 ~{\pm}~ 0.13$   &   77.19	\\
{Q0329-3850} &     $0.970957$   &     $0.051 ~{\pm}~ 0.001$   &    $0.031 ~{\pm}~ 0.002$   &    $1.65 ~{\pm}~ 0.11$   &   76.23	\\
{Q0329-2550} &     $0.992631$   &     $0.283 ~{\pm}~ 0.011$   &    $0.167 ~{\pm}~ 0.006$   &    $1.69 ~{\pm}~ 0.09$   &   77.8	\\
{Q1448-232} &     $1.019089$   &     $0.033 ~{\pm}~ 0.005$   &    $0.015 ~{\pm}~ 0.002$   &    $2.20 ~{\pm}~ 0.44$   &   73.37	\\
{Q0453-4230} &     $1.039514$   &     $0.189 ~{\pm}~ 0.003$   &    $0.096 ~{\pm}~ 0.001$   &    $1.97 ~{\pm}~ 0.04$   &   77.61	\\
{Q2217-2818} &     $1.054310$   &     $0.046 ~{\pm}~ 0.002$   &    $0.024 ~{\pm}~ 0.001$   &    $1.92 ~{\pm}~ 0.12$   &   75.88	\\
{Q2217-2818} &     $1.082920$   &     $0.125 ~{\pm}~ 0.001$   &    $0.064 ~{\pm}~ 0.001$   &    $1.95 ~{\pm}~ 0.03$   &   77.34	\\
{Q0042-2930} &     $1.091866$   &     $0.162 ~{\pm}~ 0.005$   &    $0.136 ~{\pm}~ 0.004$   &    $1.19 ~{\pm}~ 0.05$   &   77.53	\\
{Q0926-0201} &     $1.096336$   &     $0.020 ~{\pm}~ 0.001$   &    $0.015 ~{\pm}~ 0.001$   &    $1.33 ~{\pm}~ 0.11$   &   66.11	\\
{Q2222-3939} &     $1.098126$   &     $0.184 ~{\pm}~ 0.015$   &    $0.136 ~{\pm}~ 0.015$   &    $1.35 ~{\pm}~ 0.19$   &   77.61	\\
{Q1444+014} &     $1.102026$   &     $0.142 ~{\pm}~ 0.001$   &    $0.181 ~{\pm}~ 0.001$   &    $0.78 ~{\pm}~ 0.01$   &   77.47	\\
{Q2347-4342} &     $1.109640$   &     $0.040 ~{\pm}~ 0.004$   &    $0.028 ~{\pm}~ 0.003$   &    $1.43 ~{\pm}~ 0.21$   &   75.51	\\
{Q1444+014} &     $1.129162$   &     $0.244 ~{\pm}~ 0.006$   &    $0.151 ~{\pm}~ 0.007$   &    $1.62 ~{\pm}~ 0.08$   &   77.72	\\
{Q0013-0029} &     $1.146810$   &     $0.047 ~{\pm}~ 0.001$   &    $0.019 ~{\pm}~ 0.001$   &    $2.47 ~{\pm}~ 0.14$   &   75.33	\\
{Q1151+068} &     $1.153727$   &     $0.108 ~{\pm}~ 0.003$   &    $0.077 ~{\pm}~ 0.003$   &    $1.40 ~{\pm}~ 0.07$   &   77.35	\\
{CTQ0298} &     $1.160456$   &     $0.049 ~{\pm}~ 0.003$   &    $0.033 ~{\pm}~ 0.003$   &    $1.48 ~{\pm}~ 0.16$   &   76.16	\\
{Q1621-0042} &     $1.174581$   &     $0.237 ~{\pm}~ 0.012$   &    $0.115 ~{\pm}~ 0.025$   &    $2.06 ~{\pm}~ 0.45$   &   77.66	\\
{Q0109-3518} &     $1.182684$   &     $0.135 ~{\pm}~ 0.001$   &    $0.098 ~{\pm}~ 0.001$   &    $1.38 ~{\pm}~ 0.02$   &   77.42	\\
{Q0237-23} &     $1.184633$   &     $0.140 ~{\pm}~ 0.005$   &    $0.083 ~{\pm}~ 0.004$   &    $1.69 ~{\pm}~ 0.10$   &   77.41	\\
{Q2217-2818} &     $1.200162$	&     $0.099 ~{\pm}~ 0.002$	&    $0.043 ~{\pm}~  0.001$	&    $2.30 ~{\pm}~ 0.07$	&   77.24	\\
{Q0421-2624} &     $1.210051$   &     $0.065 ~{\pm}~ 0.002$   &    $0.032 ~{\pm}~ 0.002$   &    $2.03 ~{\pm}~ 0.14$   &   76.60	\\
{Q2222-3939} &     $1.227553$   &     $0.114 ~{\pm}~ 0.005$   &    $0.042 ~{\pm}~ 0.003$   &    $2.71 ~{\pm}~ 0.23$   &   77.14	\\
{Q0926-0201} &     $1.232203$   &     $0.069 ~{\pm}~ 0.004$   &    $0.032 ~{\pm}~ 0.003$   &    $2.16 ~{\pm}~ 0.24$   &   76.64	\\
{Q1122-1648} &     $1.234160$   &     $0.200 ~{\pm}~ 0.000$   &    $0.130 ~{\pm}~ 0.002$   &    $1.54 ~{\pm}~ 0.02$   &   77.63	\\
{Q2059-360} &     $1.242973$   &     $0.015 ~{\pm}~ 0.001$   &    $0.010 ~{\pm}~ 0.002$   &    $1.50 ~{\pm}~ 0.32$   &   58.21	\\
{Q2000-330} &     $1.249864$   &     $0.032 ~{\pm}~ 0.001$   &    $0.022 ~{\pm}~ 0.001$   &    $1.45 ~{\pm}~ 0.08$   &   74.13	\\
{CTQ0298} &     $1.256069$   &     $0.057 ~{\pm}~ 0.004$   &    $0.068 ~{\pm}~ 0.002$   &    $0.84 ~{\pm}~ 0.06$   &   76.64	\\
{Q0136-231} &     $1.261761$   &     $0.102 ~{\pm}~ 0.003$   &    $0.071 ~{\pm}~ 0.005$   &    $1.44 ~{\pm}~ 0.11$   &   77.33	\\
{Q1209+0919} &     $1.264983$   &     $0.083 ~{\pm}~ 0.007$   &    $0.061 ~{\pm}~ 0.007$   &    $1.35 ~{\pm}~ 0.19$   &   77.14	\\
{Q0328-272} &     $1.269054$   &     $0.047 ~{\pm}~ 0.011$   &    $0.043 ~{\pm}~ 0.085$   &    $1.09 ~{\pm}~ 2.17$   &   76.13	\\
{Q0136-231} &     $1.285796$   &     $0.021 ~{\pm}~ 0.003$   &    $0.015 ~{\pm}~ 0.003$   &    $1.40 ~{\pm}~ 0.34$   &   67.58	\\
{Q2206-199} &     $1.297044$   &     $0.148 ~{\pm}~ 0.001$   &    $0.130 ~{\pm}~ 0.001$   &    $1.14 ~{\pm}~ 0.01$   &   77.50	\\
{Q1157+014} &     $1.330502$   &     $0.120 ~{\pm}~ 0.002$   &    $0.075 ~{\pm}~ 0.003$   &    $1.60 ~{\pm}~ 0.07$   &   77.36	\\
{Q2204-408} &     $1.335251$   &     $0.052 ~{\pm}~ 0.004$   &    $0.040 ~{\pm}~ 0.004$   &    $1.30 ~{\pm}~ 0.16$   &   76.37	\\
{Q2044-168} &     $1.342492$   &     $0.057 ~{\pm}~ 0.004$   &    $0.035 ~{\pm}~ 0.004$   &    $1.63 ~{\pm}~ 0.22$   &   76.52	\\
{Q2000-330} &     $1.342771$   &     $0.032 ~{\pm}~ 0.002$   &    $0.015 ~{\pm}~ 0.001$   &    $2.13 ~{\pm}~ 0.19$   &   73.25	\\
{Q0549-213} &     $1.343495$   &     $0.181 ~{\pm}~ 0.010$   &    $0.086 ~{\pm}~ 0.005$   &    $2.10 ~{\pm}~ 0.17$   &   77.54	\\
{Q0136-231} &     $1.353687$   &     $0.170 ~{\pm}~ 0.004$   &    $0.110 ~{\pm}~ 0.007$   &    $1.55 ~{\pm}~ 0.10$   &   77.57	\\
{Q1629+120} &     $1.379330$   &     $0.142 ~{\pm}~ 0.007$   &    $0.093 ~{\pm}~ 0.012$   &    $1.53 ~{\pm}~ 0.21$   &   77.44	\\
{Q2243-6031} &     $1.389597$   &     $0.106 ~{\pm}~ 0.022$   &    $0.054 ~{\pm}~ 0.039$   &    $1.96 ~{\pm}~ 1.48$   &   77.27	\\
{Q0011+0055} &     $1.395656$   &     $0.186 ~{\pm}~ 0.004$   &    $0.175 ~{\pm}~ 0.007$   &    $1.06 ~{\pm}~ 0.05$   &   77.63	\\
{Q0128-2150} &     $1.398315$   &     $0.018 ~{\pm}~ 0.001$   &    $0.015 ~{\pm}~ 0.002$   &    $1.20 ~{\pm}~ 0.17$   &   63.24	\\
{Q0951-0450} &     $1.399375$   &     $0.073 ~{\pm}~ 0.004$   &    $0.041 ~{\pm}~ 0.010$   &    $1.78 ~{\pm}~ 0.45$   &   76.86	\\
{Q2059-360} &     $1.399911$   &     $0.109 ~{\pm}~ 0.002$   &    $0.061 ~{\pm}~ 0.003$   &    $1.79 ~{\pm}~ 0.09$   &   77.31	\\
{Q2347-4342} &     $1.405367$   &     $0.074 ~{\pm}~ 0.001$   &    $0.041 ~{\pm}~ 0.001$   &    $1.80 ~{\pm}~ 0.05$   &   76.86	\\
{Q2225-2258} &     $1.412608$   &     $0.271 ~{\pm}~ 0.002$   &    $0.150 ~{\pm}~ 0.014$   &    $1.81 ~{\pm}~ 0.17$   &   77.76	\\
{Q0128-2150} &     $1.422159$   &     $0.042 ~{\pm}~ 0.001$   &    $0.020 ~{\pm}~ 0.001$   &    $2.10 ~{\pm}~ 0.12$   &   75.36	\\
{Q2225-2258} &     $1.432967$   &     $0.167 ~{\pm}~ 0.002$   &    $0.083 ~{\pm}~ 0.001$   &    $2.01 ~{\pm}~ 0.03$   &   77.48	\\
{Q0002-4220} &     $1.446496$   &     $0.042 ~{\pm}~ 0.000$   &    $0.026 ~{\pm}~ 0.000$   &    $1.62 ~{\pm}~ 0.00$   &   75.73	\\
{Q0122-380} &     $1.449964$   &     $0.061 ~{\pm}~ 0.006$   &    $0.044 ~{\pm}~ 0.023$   &    $1.39 ~{\pm}~ 0.74$   &   76.70	\\
{Q1448-232} &     $1.473252$   &     $0.269 ~{\pm}~ 0.009$   &    $0.198 ~{\pm}~ 0.008$   &    $1.36 ~{\pm}~ 0.07$   &   77.78	\\
{Q0551-3637} &     $1.491767$   &     $0.176 ~{\pm}~ 0.004$   &    $0.087 ~{\pm}~ 0.003$   &    $2.02 ~{\pm}~ 0.08$   &   77.53	\\
{Q1418-064} &     $1.516673$   &     $0.075 ~{\pm}~ 0.003$   &    $0.047 ~{\pm}~ 0.003$   &    $1.60 ~{\pm}~ 0.12$   &   76.93	\\
{Q2217-2818} &     $1.555884$   &     $0.268 ~{\pm}~ 0.001$   &    $0.182 ~{\pm}~ 0.001$   &    $1.47 ~{\pm}~ 0.01$   &   77.77	\\
{Q1448-232} &     $1.585464$   &     $0.075 ~{\pm}~ 0.001$   &    $0.056 ~{\pm}~ 0.001$   &    $1.34 ~{\pm}~ 0.03$   &   76.96	\\
{Q2225-2258} &     $1.639427$   &     $0.277 ~{\pm}~ 0.002$   &    $0.214 ~{\pm}~ 0.003$   &    $1.29 ~{\pm}~ 0.02$   &   77.81	\\
{Q0001-2340} &     $1.651484$   &     $0.068 ~{\pm}~ 0.001$   &    $0.046 ~{\pm}~ 0.001$   &    $1.48 ~{\pm}~ 0.04$   &   76.84	\\
{Q0429-4901} &     $1.680766$   &     $0.023 ~{\pm}~ 0.001$   &    $0.010 ~{\pm}~ 0.001$   &    $2.30 ~{\pm}~ 0.25$   &   65.75	\\
{Q0151-4326} &     $1.708492$   &     $0.026 ~{\pm}~ 0.001$   &    $0.013 ~{\pm}~ 0.001$   &    $2.00 ~{\pm}~ 0.17$   &   70.96	\\
{Q2243-6031} &     $1.755699$   &     $0.108 ~{\pm}~ 0.001$   &    $0.057 ~{\pm}~ 0.001$   &    $1.89 ~{\pm}~ 0.04$   &   77.28	\\
{Q0100+130} &     $1.758442$   &     $0.028 ~{\pm}~ 0.004$   &    $0.016 ~{\pm}~ 0.002$   &    $1.75 ~{\pm}~ 0.33$   &   72.69	\\
{Q0011+0055} &     $1.777926$   &     $0.127 ~{\pm}~ 0.003$   &    $0.084 ~{\pm}~ 0.003$   &    $1.51 ~{\pm}~ 0.06$   &   77.36	\\
{Q0141-3932} &     $1.781686$   &     $0.042 ~{\pm}~ 0.001$   &    $0.024 ~{\pm}~ 0.001$   &    $1.75 ~{\pm}~ 0.08$   &   75.67	\\
{Q2347-4342} &     $1.796233$   &     $0.147 ~{\pm}~ 0.001$   &    $0.120 ~{\pm}~ 0.001$   &    $1.23 ~{\pm}~ 0.01$   &   77.50	\\
{Q0453-4230} &     $1.858369$   &     $0.194 ~{\pm}~ 0.001$   &    $0.149 ~{\pm}~ 0.001$   &    $1.30 ~{\pm}~ 0.01$   &   77.63	\\
{Q1418-064} &     $1.883599$   &     $0.017 ~{\pm}~ 0.003$   &    $0.016 ~{\pm}~ 0.004$   &    $1.06 ~{\pm}~ 0.33$   &   61.83	\\
{Q0122-380} &     $1.911032$   &     $0.158 ~{\pm}~ 0.002$   &    $0.104 ~{\pm}~ 0.002$   &    $1.52 ~{\pm}~ 0.03$   &   77.52	\\
{Q0122-380} &     $1.974115$   &     $0.279 ~{\pm}~ 0.051$   &    $0.181 ~{\pm}~ 0.026$   &    $1.54 ~{\pm}~ 0.36$   &   77.80	\\
{Q0002-4220} &     $1.988641$   &     $0.285 ~{\pm}~ 0.001$   &    $0.212 ~{\pm}~ 0.002$   &    $1.34 ~{\pm}~ 0.01$   &   77.81	\\
{Q1341-1020} &     $2.147324$   &     $0.289 ~{\pm}~ 0.009$   &    $0.205 ~{\pm}~ 0.067$   &    $1.41 ~{\pm}~ 0.46$   &   77.81	\\
{Q1418-064} &     $2.174224$   &     $0.178 ~{\pm}~ 0.004$   &    $0.122 ~{\pm}~ 0.004$   &    $1.46 ~{\pm}~ 0.06$   &   77.58	\\
{Q0940-1050} &     $2.174535$   &     $0.028 ~{\pm}~ 0.001$   &    $0.026 ~{\pm}~ 0.001$   &    $1.08 ~{\pm}~ 0.06$   &   72.98	\\
{Q1140+2711} &     $2.196649$   &     $0.193 ~{\pm}~ 0.002$   &    $0.125 ~{\pm}~ 0.002$   &    $1.54 ~{\pm}~ 0.03$   &   77.63	\\
{Q0100+130} &     $2.298051$   &     $0.230 ~{\pm}~ 0.004$   &    $0.154 ~{\pm}~ 0.004$   &    $1.49 ~{\pm}~ 0.05$         &   77.72	\\

\enddata
\label{tab:2}
\tablecomments{\scriptsize{This table lists the details of the $116$ weak {\MgII} systems detected in our sample of $81$ quasars. The second column refers to the redshift of the absorber. The third and fourth columns are the measured rest-frame equivalent widths of {\MgII}~$\lambda 2796$~{\AA} and {\MgII}~$\lambda 2803$~{\AA} respectively. The fifth column is the doublet ratio given by $W_r(2796)/W_r(2803)$. The final column is the tcumulative redshift path length for each system. The first four listed systems have $z < 0.4$, and are therefore not part of our survey.}}
\end{deluxetable}

\clearpage

\begin{deluxetable}{lcccc}
\tablenum{3}
\tablewidth{0pt}
\tablecaption{Number of absorbers per unit redshift ($dN/dz$)}
\tablehead{
\colhead{ } &
\colhead{$0.4 < z < 0.7$} &
\colhead{$0.7 < z < 1.0$} &
\colhead{$1.0 < z < 1.4$} &
\colhead{$1.4 < z < 2.4$}
}
\startdata
{CRCV99} & $1.43 \pm 0.21$ & $1.84 \pm 0.26$ & $2.19 \pm 0.80$ & $--$ \\
{LCT06}  & $1.44 \pm 0.52$ & $3.56 \pm 0.74$ & $1.44 \pm 0.23$ & $0.99 \pm 0.11$ \\
{This Survey} & $1.06 \pm 0.10$ & $1.51 \pm 0.09$ & $1.76 \pm 0.08$ & $1.06 \pm 0.04$ \\
\enddata
\label{tab:3}
\tablecomments{\small{The various $dN/dz$ values estimated in this and previous surveys. CRCV99 refers to the \citet{weak1} Keck/HIRES survey of weak {\MgII} absorbers over the redshift interval $0.4 < z < 1.4$. LCT06 refers to the \citet{LCT06a} VLT/UVES survey over the redshift interval $0.4 < z < 1.4$. These values are plotted in Figure~\ref{fig:4}.}}
\end{deluxetable}

\end{document}